\documentclass[5p,times]{cas-dc}

\usepackage[square,sort,comma,numbers]{natbib}

\usepackage[cmex10]{amsmath}

\usepackage{array}
\usepackage{mdwmath}
\usepackage{mdwtab}
\usepackage{eqparbox}
\usepackage{url}
\usepackage{supertabular}                                             
\usepackage{amssymb}
\usepackage{caption}
\usepackage{subcaption}
\usepackage{float}
\usepackage{booktabs}
\usepackage{multirow}
\usepackage{graphicx}
\usepackage{color}
\usepackage{tabularray}
\usepackage{hhline}
\usepackage{wasysym}

\usepackage[linesnumbered,ruled,vlined]{algorithm2e}
\usepackage{amsmath}
\usepackage{amssymb}

\usepackage{float}
\usepackage{placeins}
\usepackage{subcaption}

\usepackage{multirow}
\usepackage{float}
\usepackage{afterpage}
\usepackage{placeins}

\setcitestyle{square}
\def\tsc#1{\csdef{#1}{\textsc{\lowercase{#1}}\xspace}}
\tsc{WGM}
\tsc{QE}
\tsc{EP}
\tsc{PMS}
\tsc{BEC}
\tsc{DE}

\begin{document}
\let\WriteBookmarks\relax
\def\floatpagepagefraction{1}
\def\textpagefraction{.001}

\shorttitle{ReinFog: A DRL Empowered Framework for Resource Management in Edge and Cloud Computing Environments}

\shortauthors{Wang, Goudarzi, Buyya}

\title [mode = title]{ReinFog: A Deep Reinforcement Learning Empowered Framework for Resource Management in Edge and Cloud Computing Environments}                      

\author[1]{Zhiyu Wang}
\ead{zhiywang1@student.unimelb.edu.au}

\affiliation[1]{organization={The Quantum Cloud Computing and Distributed Systems (qCLOUDS) Laboratory, School of Computing and Information Systems, The University of Melbourne},
    city={Melbourne},
    country={Australia}}

\author[2]{Mohammad Goudarzi}

\affiliation[2]{organization={The Faculty of Information Technology, Monash University},
    city={Melbourne},
    country={Australia}}
\ead{mohammad.goudarzi@monash.edu}

\author[1]{Rajkumar Buyya}
\ead{rbuyya@unimelb.edu.au}

\begin{abstract}
The growing IoT landscape requires effective server deployment strategies to meet demands including real-time processing and energy efficiency. This is complicated by heterogeneous, dynamic applications and servers. To address these challenges, we propose ReinFog, a modular distributed software empowered with Deep Reinforcement Learning (DRL) for adaptive resource management across edge/fog and cloud environments. ReinFog enables the practical development/deployment of various centralized and distributed DRL techniques for resource management in edge/fog and cloud computing environments. It also supports integrating native and library-based DRL techniques for diverse IoT application scheduling objectives. Additionally, ReinFog allows for customizing deployment configurations for different DRL techniques, including the number and placement of DRL Learners and DRL Workers in large-scale distributed systems. Besides, we propose a novel Memetic Algorithm for DRL Component (e.g., DRL Learners and DRL Workers) Placement in ReinFog named MADCP, which combines the strengths of Genetic Algorithm, Firefly Algorithm, and Particle Swarm Optimization. Experiments reveal that the DRL mechanisms developed within ReinFog have significantly enhanced both centralized and distributed DRL techniques implementation. These advancements have resulted in notable improvements in IoT application performance, reducing response time by 45\%, energy consumption by 39\%, and weighted cost by 37\%, while maintaining minimal scheduling overhead. Additionally, ReinFog exhibits remarkable scalability, with a rise in DRL Workers from 1 to 30 causing only a 0.3-second increase in startup time and around 2 MB more RAM per Worker. The proposed MADCP for DRL component placement further accelerates the convergence rate of DRL techniques by up to 38\%.
\end{abstract}


\maketitle

\section{Introduction}
The evolution of computing paradigms has led to the emergence of edge/fog and cloud computing as complementary approaches to address the growing demands of modern applications \cite{jamil2022resource}. Edge/fog computing refers to the deployment of computational resources closer to data sources, such as IoT devices, sensors, and gateways, to reduce latency and improve real-time processing capabilities \cite{meruje2024databases}. This distributed network of nodes performs data processing, storage, and decision-making near the data origin, minimizing the need to send large volumes of data to distant servers. By processing tasks locally or in proximity, edge/fog computing helps to reduce latency and enhance the responsiveness of time-sensitive applications. In contrast, cloud computing offers centralized resources in remote data centers, providing vast storage and processing capabilities ideal for handling large-scale data and complex computations, albeit with potentially higher latency for real-time applications \cite{mansouri2021review}.

The synergy between edge/fog and cloud computing creates a powerful infrastructure capable of supporting various applications with varying requirements. This integrated approach allows critical applications to be processed closer to the data source for faster response times, while less time-sensitive and resource-intensive applications can be offloaded to the cloud. Such a hybrid model is particularly crucial in the context of the Internet of Things (IoT), where applications are growing at an unprecedented rate \cite{jeyaraj2023resource}. This has led to a significant increase in data generation and processing demands, necessitating efficient deployment strategies that can effectively leverage the computational capabilities across edge/fog and cloud environments \cite{jin2023cloud}.

As IoT applications continue to evolve and expand, they present unique challenges for resource management and scheduling \cite{goudarzi2022scheduling}. The heterogeneity of IoT devices and edge/fog and cloud environments, ranging from resource-constrained sensors to powerful cloud servers, is characterized by varying hardware capabilities (e.g., processing power, memory, and energy constraints), network conditions (e.g., latency, bandwidth, and reliability), and application requirements (e.g., real-time processing versus compute-intensive tasks) \cite{buyya2023quality}. This diverse landscape makes rule-based resource scheduling strategies ineffective \cite{10316602}. Moreover, the dynamic nature of IoT workloads and network conditions requires adaptive resource management solutions capable of responding to rapid and unpredictable changes \cite{sharif2024hybrid}. The exponential growth in the number of IoT devices and applications further compounds these issues, demanding highly scalable management approaches. Additionally, IoT applications often have conflicting requirements, such as minimizing latency while maximizing energy efficiency \cite{ali2023systematic}, necessitating complex multi-objective optimization strategies.

Traditional heuristic or rule-based approaches to resource management often rely on static optimization or pre-defined rules, which can be effective in predictable environments but fall short in adapting to the rapidly changing conditions of IoT ecosystems \cite{chen2021deep}. These methods struggle to optimize multiple objectives such as minimizing response time and reducing energy consumption \cite{wang2024deep}, particularly as the complexity of decision-making increases exponentially with the scale of the system \cite{huang2024reinforcement}. For instance, in real-time video processing applications deployed across edge/fog and cloud nodes, the system must continuously adapt to fluctuating bandwidth, latency, and processing power  \cite{wang2022container}. Heuristic methods may struggle to balance these dynamic trade-offs, often resulting in suboptimal resource scheduling and increased system latency. In contrast, Deep Reinforcement Learning (DRL) techniques offer a more adaptive and scalable solution by continuously learning and optimizing resource management decisions based on real-time feedback from the environment \cite{zhou2022deep}. This enables DRL-based approaches to dynamically schedule IoT applications, predict future workloads, and efficiently utilize available resources, outperforming traditional methods in complex and unpredictable IoT environments.

While DRL techniques demonstrate significant potential in addressing dynamic resource management challenges, to the best of our knowledge, there is currently no framework that comprehensively integrates both centralized and distributed DRL techniques for IoT application scheduling in edge/fog and cloud computing environments. Existing frameworks primarily rely on rule-based and heuristic methods. This critical gap is reflected in two key aspects. First, current solutions do not provide mechanisms to simultaneously accommodate centralized and distributed DRL techniques, which is essential for efficient resource management across dynamic and stochastic edge/fog and cloud environments. Second, existing frameworks cannot support both native DRL technique implementations and external DRL library integrations, limiting the flexibility and adaptability of DRL-based solutions in heterogeneous computing environments. These limitations highlight the urgent need for a unified framework that can effectively leverage various DRL techniques for IoT application scheduling in edge/fog and cloud computing environments.

To address these challenges, we propose ReinFog, a novel framework that harnesses the power of DRL for adaptive resource management in edge/fog and cloud computing environments. To the best of our knowledge, ReinFog is the first framework that comprehensively integrates mechanisms for the integration of both centralized and distributed DRL techniques for IoT application scheduling, while supporting both native DRL implementations and external DRL library integrations through a modular and extensible design. To address the dynamic nature of IoT ecosystems, ReinFog incorporates multiple DRL techniques, both centralized and distributed, to adapt to rapidly changing workloads and network conditions, enabling real-time, intelligent IoT application scheduling decisions that optimize multiple objectives simultaneously. ReinFog supports native DRL technique implementations, enabling researchers to design and develop centralized and distributed DRL techniques, specifically designed for edge/fog and cloud computing environments. Also, recognizing that each DRL library incorporates a set of specific DRL techniques, ReinFog offers mechanisms to seamlessly integrate external DRL libraries.  This dual capability enhances the flexibility and adaptability of DRL-based solutions in heterogeneous computing environments, allowing users to leverage familiar tools and accelerate their development processes. Accordingly, to facilitate the implementation of both native and library-based DRL techniques, ReinFog adopts a modular design that supports easy extension and customization. This design separates DRL components into DRL Workers for environment interaction and DRL Learners for policy optimization. To accommodate the diverse requirements of large-scale distributed systems, ReinFog supports customizable deployment configurations for DRL techniques. This design allows for flexible configuration of DRL Learners and DRL Workers, enabling users to tailor deployments according to specific system architectures and performance needs. Such flexibility is crucial for effectively managing resources and ensuring optimal performance in complex IoT environments with varying scales and topologies. Moreover, as the efficient execution of different DRL techniques requires interaction among multiple DRL-related components, it is crucial to place these components on appropriate nodes. To optimize the DRL component placement, we propose a novel Memetic Algorithm for DRL Component Placement in ReinFog, named MADCP. MADCP combines the Genetic Algorithm (GA)'s robust exploration capabilities, Firefly Algorithm (FA)'s ability to fine-tune local search, and Particle Swarm Optimization (PSO)'s efficient global optimization to efficiently place DRL components across heterogeneous computing nodes. This algorithm enhances ReinFog's ability to quickly adapt to changing environmental conditions and optimize resource utilization before the start of DRL training processes.

The key contributions of our paper are as follows:
\begin{itemize}
\item We propose ReinFog, a containerized and modular framework for DRL-based resource management in edge/fog and cloud environments. It offers mechanisms to support both centralized and distributed DRL techniques. Also, it enables the integration of both native DRL techniques and external DRL libraries. 
\item We design customizable deployment configurations for DRL techniques in ReinFog, allowing flexible configuration of DRL Learners and DRL Workers in large-scale distributed systems.
\item We propose a novel Memetic Algorithm for DRL Component Placement in ReinFog, named MADCP, combining GA, FA, and PSO for efficient DRL component placement.
\item We conduct extensive practical experiments evaluating ReinFog's performance across various aspects. It demonstrates that ReinFog is a lightweight and scalable framework capable of effectively scheduling IoT applications under diverse optimization objectives.
\end{itemize}

\section{Related Work}
IoT application scheduling and resource management in edge, fog, and cloud environments have attracted significant research attention. Existing approaches can be broadly categorized into two groups: (i) algorithmic techniques that focus on optimizing scheduling decisions using heuristics, meta-heuristics, or DRL, and (ii) system-level software frameworks that aim to support the practical deployment and management of IoT applications. In this section, we first review algorithmic techniques, including both heuristic/meta-heuristic and DRL-based methods. Then, we summarize relevant software frameworks and highlight their limitations in handling dynamic, large-scale, and heterogeneous environments. Finally, we provide a comparative analysis to clearly position the originality and technical contributions of our proposed ReinFog framework.

\subsection{Algorithmic techniques for IoT Scheduling}
A wide range of algorithmic techniques have been proposed to address the challenges of scheduling and resource management in IoT-enabled edge, fog, and cloud environments. These techniques can be broadly classified into heuristic/meta-heuristic methods and machine learning-based techniques. In terms of heuristic techniques, Wu et al. \cite{wu2018hybrid} modeled IoT application scheduling in edge and fog environments as a Directed Acyclic Graph (DAG), using EDA and partitioning to queue IoT applications and assign servers. Ali et al. \cite{ali2020automated} proposed an NSGA2-based technique for minimizing the total computation time and system cost of IoT application scheduling in heterogeneous fog cloud computing environments. Hoseiny et al. \cite{hoseiny2021pga} proposed a GA-based technique to minimize computation time and energy consumption in heterogeneous fog-cloud IoT application scheduling. While these heuristic methods perform well in specific scenarios, they often lack adaptability to dynamic environments. In recent years, machine learning techniques, particularly DRL, have gained significant attention for resource management in edge/fog and cloud computing environments, owing to their adaptability and capacity for continuous learning in dynamic scenarios. Huang et al. \cite{huang2019deep} applied a Deep Q-Network (DQN)-based approach to address resource allocation problems within edge computing environments. Zheng et al. \cite{zheng2022sac} proposed a Soft Actor-Critic (SAC)-based technique to solve the optimization problem of computational offloading and resource allocation in collaborative vehicle networks. Siyadatzadeh et al. \cite{siyadatzadeh2023relief} proposed ReLIEF, which employs Q-Learning to manage resources in fog-based IoT systems. Wang et al. \cite{wang2024deep} proposed DRLIS, which leverages Proximal Policy Optimization (PPO) to optimize system load balancing and response time in edge and fog computing environments. Zou et al. \cite{zou2020a3c} proposed A3C-DO, which utilizes the Asynchronous Advantage Actor-Critic (A3C) technique to manage resources in edge computing environments. Liu et al. \cite{liu2023asynchronous} proposed an A3C-based approach for edge computing in the smart vehicles domain. Wang et al. \cite{wang2025tf} proposed TF-DDRL, which is based on Importance Weighted Actor-Learner Architectures (IMPALA) to schedule IoT applications under three optimization objectives. Table \ref{tab:techcompare} presents a qualitative analysis of existing techniques proposed for IoT application scheduling. While prior studies focus on DRL-based scheduling, none provide a unified software framework that facilitates the implementation of both centralized and distributed DRL techniques, which is essential for experimental and practical deployment. These limitations are overcome by our proposed ReinFog software framework.
\renewcommand{\arraystretch}{1.5}
\begin{table*}[pos=t]
\centering
\caption{A qualitative analysis of existing techniques for IoT application scheduling}
\label{tab:techcompare}
\resizebox{\textwidth}{!}{%
\begin{tabular}{cccccccclllc}
\hline
\multicolumn{1}{|c|}{\multirow{3}{*}{Work}}         & \multicolumn{4}{c|}{Architectural Properties}                                                                                                                 & \multicolumn{6}{c|}{Algorithm Properties}                                                                                                                                 & \multicolumn{1}{c|}{\multirow{3}{*}{Evaluation}} \\ \cline{2-11}
\multicolumn{1}{|c|}{}                              & \multicolumn{2}{c|}{IoT Device Layer}                                       & \multicolumn{2}{c|}{Edge/Cloud Layer}                                           & \multicolumn{2}{c|}{\multirow{2}{*}{Main Technique}}                                             & \multicolumn{4}{c|}{\multirow{2}{*}{Multiple Optimization Objectives}} & \multicolumn{1}{c|}{}                            \\ \cline{2-5}
\multicolumn{1}{|c|}{}                              & \multicolumn{1}{c|}{Real Applications} & \multicolumn{1}{c|}{Request Type}  & \multicolumn{1}{c|}{Computing Environemnt} & \multicolumn{1}{c|}{Heterogeneity} & \multicolumn{2}{c|}{}                                                                            & \multicolumn{4}{c|}{}                                                  & \multicolumn{1}{c|}{}                            \\ \hline
\multicolumn{1}{|c|}{\cite{wu2018hybrid}}           & \multicolumn{1}{c|}{\LEFTcircle}       & \multicolumn{1}{c|}{Homogeneous}   & \multicolumn{1}{c|}{Edge and Cloud}        & \multicolumn{1}{c|}{Heterogeneous} & \multicolumn{1}{c|}{\multirow{3}{*}{Metaheuristic Algorithms}} & \multicolumn{1}{c|}{EDA}        & \multicolumn{4}{c|}{\checkmark}                                        & \multicolumn{1}{c|}{Simulation}                  \\ \cline{1-5} \cline{7-12} 
\multicolumn{1}{|c|}{\cite{ali2020automated}}       & \multicolumn{1}{c|}{\Circle}           & \multicolumn{1}{c|}{Homogeneous}   & \multicolumn{1}{c|}{Edge and Cloud}        & \multicolumn{1}{c|}{Heterogeneous} & \multicolumn{1}{c|}{}                                          & \multicolumn{1}{c|}{NSGA2}      & \multicolumn{4}{c|}{\checkmark}                                        & \multicolumn{1}{c|}{Simulation}                  \\ \cline{1-5} \cline{7-12} 
\multicolumn{1}{|c|}{\cite{hoseiny2021pga}}         & \multicolumn{1}{c|}{\Circle}           & \multicolumn{1}{c|}{Homogeneous}   & \multicolumn{1}{c|}{Edge and Cloud}        & \multicolumn{1}{c|}{Heterogeneous} & \multicolumn{1}{c|}{}                                          & \multicolumn{1}{c|}{GA}         & \multicolumn{4}{c|}{$\times$}                                          & \multicolumn{1}{c|}{Simulation}                  \\ \hline
\multicolumn{1}{|c|}{\cite{huang2019deep}}          & \multicolumn{1}{c|}{\LEFTcircle}       & \multicolumn{1}{c|}{Heterogeneous} & \multicolumn{1}{c|}{Edge}                  & \multicolumn{1}{c|}{Homogeneous}   & \multicolumn{1}{c|}{\multirow{4}{*}{Centrlized DRL}}           & \multicolumn{1}{c|}{DQN}        & \multicolumn{4}{c|}{\checkmark}                                        & \multicolumn{1}{c|}{Simulation}                  \\ \cline{1-5} \cline{7-12} 
\multicolumn{1}{|c|}{\cite{zheng2022sac}}           & \multicolumn{1}{c|}{\LEFTcircle}       & \multicolumn{1}{c|}{Homogeneous}   & \multicolumn{1}{c|}{Edge}                  & \multicolumn{1}{c|}{Homogeneous}   & \multicolumn{1}{c|}{}                                          & \multicolumn{1}{c|}{SAC}        & \multicolumn{4}{c|}{$\times$}                                          & \multicolumn{1}{c|}{Simulation}                  \\ \cline{1-5} \cline{7-12} 
\multicolumn{1}{|c|}{\cite{siyadatzadeh2023relief}} & \multicolumn{1}{c|}{\LEFTcircle}       & \multicolumn{1}{c|}{Heterogeneous} & \multicolumn{1}{c|}{Edge}                  & \multicolumn{1}{c|}{Heterogeneous} & \multicolumn{1}{c|}{}                                          & \multicolumn{1}{c|}{Q-Learning} & \multicolumn{4}{c|}{\checkmark}                                        & \multicolumn{1}{c|}{Simulation}                  \\ \cline{1-5} \cline{7-12} 
\multicolumn{1}{|c|}{\cite{wang2024deep}}           & \multicolumn{1}{c|}{\CIRCLE}           & \multicolumn{1}{c|}{Heterogeneous} & \multicolumn{1}{c|}{Edge and Cloud}        & \multicolumn{1}{c|}{Heterogeneous} & \multicolumn{1}{c|}{}                                          & \multicolumn{1}{c|}{PPO}        & \multicolumn{4}{c|}{\checkmark}                                        & \multicolumn{1}{c|}{Practical}                   \\ \hline
\multicolumn{1}{|c|}{\cite{zou2020a3c}}             & \multicolumn{1}{c|}{\LEFTcircle}       & \multicolumn{1}{c|}{Homogeneous}   & \multicolumn{1}{c|}{Edge}                  & \multicolumn{1}{c|}{Heterogeneous} & \multicolumn{1}{c|}{\multirow{3}{*}{Distributed DRL}}          & \multicolumn{1}{c|}{A3C}        & \multicolumn{4}{c|}{\checkmark}                                        & \multicolumn{1}{c|}{Simulation}                  \\ \cline{1-5} \cline{7-12} 
\multicolumn{1}{|c|}{\cite{liu2023asynchronous}}    & \multicolumn{1}{c|}{\Circle}           & \multicolumn{1}{c|}{Homogeneous}   & \multicolumn{1}{c|}{Edge and Cloud}        & \multicolumn{1}{c|}{Heterogeneous} & \multicolumn{1}{c|}{}                                          & \multicolumn{1}{c|}{A3C}        & \multicolumn{4}{c|}{$\times$}                                          & \multicolumn{1}{c|}{Simulation}                  \\ \cline{1-5} \cline{7-12} 
\multicolumn{1}{|c|}{\cite{wang2025tf}}             & \multicolumn{1}{c|}{\CIRCLE}           & \multicolumn{1}{c|}{Heterogeneous} & \multicolumn{1}{c|}{Edge and Cloud}        & \multicolumn{1}{c|}{Heterogeneous} & \multicolumn{1}{c|}{}                                          & \multicolumn{1}{c|}{IMPALA}     & \multicolumn{4}{c|}{\checkmark}                                        & \multicolumn{1}{c|}{Practical}                   \\ \hline
\multicolumn{12}{l}{\CIRCLE: Real IoT Application and Deployment, \LEFTcircle: Simulated IoT Application, \Circle: Random}                    \end{tabular}}
\end{table*}
\renewcommand{\arraystretch}{1}

\subsection{Frameworks for Resource Management}
Building upon these techniques, researchers have developed several frameworks for resource management in edge/fog and cloud computing environments. Many of these frameworks employ heuristic or meta-heuristic techniques for making resource management decisions. For instance, Yigitoglu et al. \cite{yigitoglu2017foggy} developed Foggy, a container-enabled framework supporting policy and rule-based scheduling of containerized IoT applications with dependent tasks. Similarly, Merlino et al. \cite{merlino2019enabling} proposed a framework allowing policy-driven vertical and horizontal task offloading. Yousefpour et al. \cite{yousefpour2019fogplan} introduced FogPlan, which employs greedy algorithms to minimize IoT application response time, while Ghosh et al. \cite{ghosh2019mobi} developed Mobi-IoST, using a probabilistic approach for IoT application scheduling. Deng et al. \cite{deng2021fogbus2} created FogBus2, introducing a GA-based IoT application scheduling technique, and Pallewatta et al. \cite{pallewatta2024microfog} proposed MicroFog, integrating multiple heuristic algorithms to enhance IoT application scheduling flexibility. In recent years, some frameworks have started to incorporate reinforcement learning techniques. Toosi et al. \cite{n2022greenfog} developed GreenFog, which combines linear programming optimization with the Multi-Armed Bandit approach for energy consumption reduction. Similarly, Nkenyereye et al. \cite{nkenyereye2023deep} proposed CEIF, which adopted Deep Q-Learning for resource management in edge computing environments.

\subsection{Summary and Technical Comparison with Existing Frameworks}
Table \ref{tab:compare} identifies the main properties of the related frameworks and compares them with ReinFog. Environment Support indicates the computing environments supported by each framework. DRL-Integrated Framework indicates whether the framework is comprehensively integrated with DRL capabilities, encompassing multiple DRL techniques and providing support for extensibility. The DRL Capabilities section is a specific contribution of ReinFog. These capabilities enable more adaptive and intelligent resource management by leveraging DRL to dynamically optimize IoT application scheduling in real time. It is further divided into three sub-categories. Mechanism indicates whether the framework supports the integration of native DRL techniques or importing external libraries. Architecture shows whether the framework supports centralized and distributed DRL techniques. Finally, DRL Component Placement shows whether the framework can automatically optimize the placement of DRL components. The Generic Capabilities section represents general requirements for frameworks in edge/fog or cloud computing environments. These are commonly expected features that any robust framework should provide to ensure flexibility, adaptability, and ease of integration across various platforms and environments, especially in addressing the challenges posed by heterogeneous systems. It is further divided into five sub-categories. Multi-platform Support shows whether the framework can operate across diverse heterogeneous hardware and software platforms. Container Support refers to the ability to use containerization technologies. Scalability, Configurability, and Extensibility assess the framework's ability to be scaled, customized, and incorporate new features. 

ReinFog offers significant advantages over related frameworks across multiple dimensions. Many existing frameworks rely on traditional heuristic (\cite{yigitoglu2017foggy}, \cite{merlino2019enabling}, \cite{yousefpour2019fogplan}, \cite{ghosh2019mobi}, \cite{pallewatta2024microfog}) or meta-heuristic (\cite{deng2021fogbus2}) techniques for IoT application scheduling, which often lack adaptability to manage dynamic and complex computing environments. Although some recent frameworks have started incorporating basic and single reinforcement learning techniques, such as Multi-Armed Bandit (\cite{n2022greenfog}) or Deep Q-Learning (\cite{nkenyereye2023deep}), these frameworks do not offer mechanisms for integration/development of centralized and distributed DRL techniques. Accordingly, these frameworks struggle with the implementation of efficient and scalable DRL techniques for large-scale deployments. To the best of our knowledge, ReinFog is the first resource management framework that enables the integration of both centralized and distributed DRL mechanisms for IoT application scheduling across edge/fog and cloud environments. These mechanisms enable the integration/development of a wide range of centralized and distributed techniques such as PPO \cite{schulman2017proximal} and IMPALA \cite{espeholt2018impala}. Besides, ReinFog offers interfaces for the integration of both native and library-based DRL techniques. Notably, ReinFog introduces DRL component placement, a novel feature to customize and optimize the placement of DRL components using multiple meta-heuristic algorithms and proposed MADCP. These comprehensive features and capabilities make ReinFog a versatile platform for researchers, enabling them to either utilize built-in DRL techniques or extend its mechanisms for various resource management scenarios in edge/fog and cloud computing environments. In terms of generic capabilities, ReinFog also offers multi-platform support, a feature lacking in several frameworks (e.g., \cite{yousefpour2019fogplan}, \cite{ghosh2019mobi}, \cite{n2022greenfog}). Besides, ReinFog is designed with scalability, configurability, and extensibility in mind, addressing limitations found in many existing frameworks (\cite{yigitoglu2017foggy}, \cite{merlino2019enabling}, \cite{yousefpour2019fogplan}, \cite{ghosh2019mobi}, \cite{n2022greenfog}, \cite{nkenyereye2023deep}).

\renewcommand{\arraystretch}{1.5}
\begin{table*}[pos=t]
\centering
\caption{A qualitative comparison of related software frameworks with ReinFog}
\label{tab:compare}
\resizebox{\textwidth}{!}{%
\begin{tabular}{ccccccccccccc}
\hline
\multicolumn{1}{|c|}{\multirow{3}{*}{Work}}         & \multicolumn{1}{c|}{\multirow{3}{*}{Environment Support}}                       & \multicolumn{1}{c|}{\multirow{3}{*}{\begin{tabular}[c]{@{}c@{}}DRL-Integrated \\ Framework\end{tabular}}} & \multicolumn{5}{c|}{DRL Capabilities}                                                                                                                                                                                                               & \multicolumn{5}{c|}{Generic Capabilities}                                                                                                                                                                                                                                                                                           \\ \cline{4-13} 
\multicolumn{1}{|c|}{}                              & \multicolumn{1}{c|}{}                                                           & \multicolumn{1}{c|}{}                                                                                     & \multicolumn{2}{c|}{Mechanism}                                    & \multicolumn{2}{c|}{Architecture}                                   & \multicolumn{1}{c|}{\multirow{2}{*}{\begin{tabular}[c]{@{}c@{}}DRL Components \\ Placement\end{tabular}}} & \multicolumn{1}{c|}{\multirow{2}{*}{\begin{tabular}[c]{@{}c@{}}Multi-platform \\ Support\end{tabular}}} & \multicolumn{1}{c|}{\multirow{2}{*}{Container Support}} & \multicolumn{1}{c|}{\multirow{2}{*}{Scalability}} & \multicolumn{1}{c|}{\multirow{2}{*}{Configurability}} & \multicolumn{1}{c|}{\multirow{2}{*}{Extensibility}} \\ \cline{4-7}
\multicolumn{1}{|c|}{}                              & \multicolumn{1}{c|}{}                                                           & \multicolumn{1}{c|}{}                                                                                     & \multicolumn{1}{c|}{Native}     & \multicolumn{1}{c|}{Library}    & \multicolumn{1}{c|}{Centralized} & \multicolumn{1}{c|}{Distributed} & \multicolumn{1}{c|}{}                                                                                     & \multicolumn{1}{c|}{}                                                                                   & \multicolumn{1}{c|}{}                                   & \multicolumn{1}{c|}{}                             & \multicolumn{1}{c|}{}                                 & \multicolumn{1}{c|}{}                               \\ \hline
\multicolumn{1}{|c|}{\cite{yigitoglu2017foggy}}     & \multicolumn{1}{c|}{\begin{tabular}[c]{@{}c@{}}Edge/Fog, \\ Cloud\end{tabular}} & \multicolumn{1}{c|}{$\times$}                                                                             & \multicolumn{1}{c|}{$\times$}   & \multicolumn{1}{c|}{$\times$}   & \multicolumn{1}{c|}{$\times$}    & \multicolumn{1}{c|}{$\times$}    & \multicolumn{1}{c|}{$\times$}                                                                             & \multicolumn{1}{c|}{\checkmark}                                                                         & \multicolumn{1}{c|}{\checkmark}                         & \multicolumn{1}{c|}{\checkmark}                   & \multicolumn{1}{c|}{$\times$}                         & \multicolumn{1}{c|}{$\times$}                       \\ \hline
\multicolumn{1}{|c|}{\cite{merlino2019enabling}}    & \multicolumn{1}{c|}{\begin{tabular}[c]{@{}c@{}}Edge/Fog, \\ Cloud\end{tabular}} & \multicolumn{1}{c|}{$\times$}                                                                             & \multicolumn{1}{c|}{$\times$}   & \multicolumn{1}{c|}{$\times$}   & \multicolumn{1}{c|}{$\times$}    & \multicolumn{1}{c|}{$\times$}    & \multicolumn{1}{c|}{$\times$}                                                                             & \multicolumn{1}{c|}{\checkmark}                                                                         & \multicolumn{1}{c|}{\checkmark}                         & \multicolumn{1}{c|}{$\triangle$}                  & \multicolumn{1}{c|}{\checkmark}                       & \multicolumn{1}{c|}{$\times$}                       \\ \hline
\multicolumn{1}{|c|}{\cite{yousefpour2019fogplan}}  & \multicolumn{1}{c|}{\begin{tabular}[c]{@{}c@{}}Edge/Fog, \\ Cloud\end{tabular}} & \multicolumn{1}{c|}{$\times$}                                                                             & \multicolumn{1}{c|}{$\times$}   & \multicolumn{1}{c|}{$\times$}   & \multicolumn{1}{c|}{$\times$}    & \multicolumn{1}{c|}{$\times$}    & \multicolumn{1}{c|}{$\times$}                                                                             & \multicolumn{1}{c|}{$\times$}                                                                           & \multicolumn{1}{c|}{\checkmark}                         & \multicolumn{1}{c|}{$\triangle$}                  & \multicolumn{1}{c|}{$\triangle$}                      & \multicolumn{1}{c|}{\checkmark}                     \\ \hline
\multicolumn{1}{|c|}{\cite{ghosh2019mobi}}          & \multicolumn{1}{c|}{\begin{tabular}[c]{@{}c@{}}Edge/Fog, \\ Cloud\end{tabular}} & \multicolumn{1}{c|}{$\times$}                                                                             & \multicolumn{1}{c|}{$\times$}   & \multicolumn{1}{c|}{$\times$}   & \multicolumn{1}{c|}{$\times$}    & \multicolumn{1}{c|}{$\times$}    & \multicolumn{1}{c|}{$\times$}                                                                             & \multicolumn{1}{c|}{$\times$}                                                                           & \multicolumn{1}{c|}{$\times$}                           & \multicolumn{1}{c|}{$\times$}                     & \multicolumn{1}{c|}{$\times$}                         & \multicolumn{1}{c|}{$\times$}                       \\ \hline
\multicolumn{1}{|c|}{\cite{deng2021fogbus2}}        & \multicolumn{1}{c|}{\begin{tabular}[c]{@{}c@{}}Edge/Fog, \\ Cloud\end{tabular}} & \multicolumn{1}{c|}{$\times$}                                                                             & \multicolumn{1}{c|}{$\times$}   & \multicolumn{1}{c|}{$\times$}   & \multicolumn{1}{c|}{$\times$}    & \multicolumn{1}{c|}{$\times$}    & \multicolumn{1}{c|}{$\times$}                                                                             & \multicolumn{1}{c|}{\checkmark}                                                                         & \multicolumn{1}{c|}{\checkmark}                         & \multicolumn{1}{c|}{\checkmark}                   & \multicolumn{1}{c|}{\checkmark}                       & \multicolumn{1}{c|}{\checkmark}                     \\ \hline
\multicolumn{1}{|c|}{\cite{pallewatta2024microfog}} & \multicolumn{1}{c|}{\begin{tabular}[c]{@{}c@{}}Edge/Fog, \\ Cloud\end{tabular}} & \multicolumn{1}{c|}{$\times$}                                                                             & \multicolumn{1}{c|}{$\times$}   & \multicolumn{1}{c|}{$\times$}   & \multicolumn{1}{c|}{$\times$}    & \multicolumn{1}{c|}{$\times$}    & \multicolumn{1}{c|}{$\times$}                                                                             & \multicolumn{1}{c|}{$\times$}                                                                         & \multicolumn{1}{c|}{\checkmark}                         & \multicolumn{1}{c|}{\checkmark}                   & \multicolumn{1}{c|}{\checkmark}                       & \multicolumn{1}{c|}{$\triangle$}                     \\ \hline
\multicolumn{1}{|c|}{\cite{n2022greenfog}}          & \multicolumn{1}{c|}{Edge/Fog}                                                   & \multicolumn{1}{c|}{$\times$}                                                                             & \multicolumn{1}{c|}{$\times$}   & \multicolumn{1}{c|}{$\times$}   & \multicolumn{1}{c|}{$\times$}    & \multicolumn{1}{c|}{$\times$}    & \multicolumn{1}{c|}{$\times$}                                                                             & \multicolumn{1}{c|}{$\times$}                                                                           & \multicolumn{1}{c|}{\checkmark}                         & \multicolumn{1}{c|}{\checkmark}                   & \multicolumn{1}{c|}{$\triangle$}                      & \multicolumn{1}{c|}{$\times$}                       \\ \hline
\multicolumn{1}{|c|}{\cite{nkenyereye2023deep}}     & \multicolumn{1}{c|}{Edge/Fog}                                                   & \multicolumn{1}{c|}{$\times$}                                                                             & \multicolumn{1}{c|}{$\times$}   & \multicolumn{1}{c|}{$\times$}   & \multicolumn{1}{c|}{$\times$}    & \multicolumn{1}{c|}{$\times$}    & \multicolumn{1}{c|}{$\times$}                                                                             & \multicolumn{1}{c|}{\checkmark}                                                                         & \multicolumn{1}{c|}{\checkmark}                         & \multicolumn{1}{c|}{\checkmark}                   & \multicolumn{1}{c|}{$\triangle$}                      & \multicolumn{1}{c|}{$\times$}                       \\ \hline
\multicolumn{1}{|c|}{ReinFog}                       & \multicolumn{1}{c|}{\begin{tabular}[c]{@{}c@{}}Edge/Fog, \\ Cloud\end{tabular}} & \multicolumn{1}{c|}{\checkmark}                                                                           & \multicolumn{1}{c|}{\checkmark} & \multicolumn{1}{c|}{\checkmark} & \multicolumn{1}{c|}{\checkmark}  & \multicolumn{1}{c|}{\checkmark}  & \multicolumn{1}{c|}{\checkmark}                                                                           & \multicolumn{1}{c|}{\checkmark}                                                                         & \multicolumn{1}{c|}{\checkmark}                         & \multicolumn{1}{c|}{\checkmark}                   & \multicolumn{1}{c|}{\checkmark}                       & \multicolumn{1}{c|}{\checkmark}                     \\ \hline
\multicolumn{13}{l}{\checkmark: Fully Supported, $\triangle$: Partially Supported, $\times$: Not Supported}     \end{tabular}
}
\end{table*}
\renewcommand{\arraystretch}{1}

\section{ReinFog Framework Architecture}
This section introduces the ReinFog framework, outlining its hardware environment and software architecture. We propose a multi-layered structure that supports heterogeneous IoT, edge/fog, and cloud environments, and detail the overall architecture of our framework.

\subsection{Hardware Environment}
\begin{figure}[pos=t]
\includegraphics[width=\linewidth]{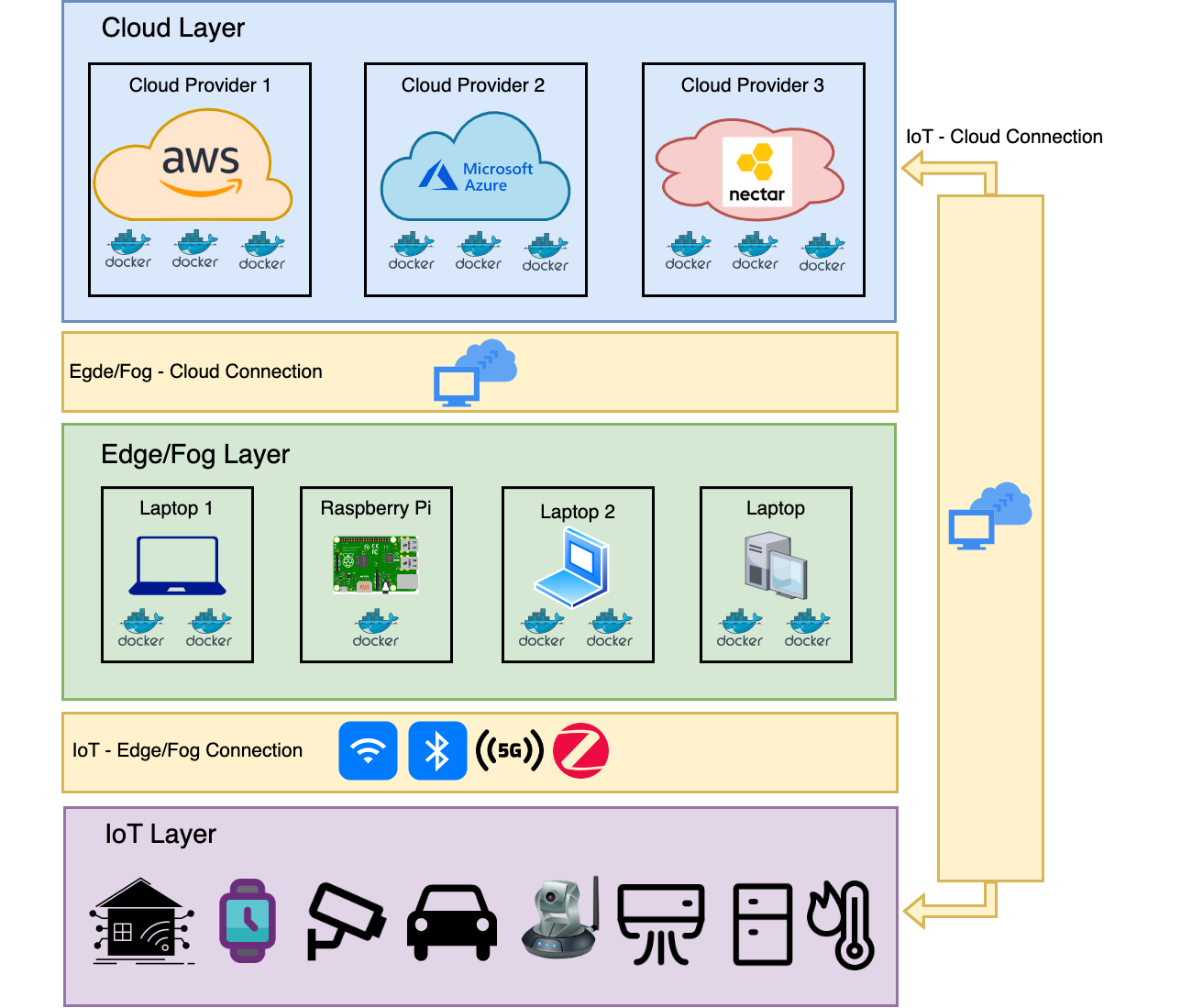}
\caption{Heterogeneous multi-layered hardware environment for ReinFog} \label{hardware}
\end{figure}

ReinFog is designed to operate across a heterogeneous multi-layered hardware environment, as illustrated in Fig. \ref{hardware}. This environment encompasses three primary layers: Cloud, Edge/Fog, and IoT, each with distinct characteristics and roles in the overall system.

\subsubsection{Cloud Layer}
The Cloud Layer represents the highest tier of computing resources in the ReinFog hardware environment. It consists of high-performance servers provided by different cloud service providers such as Amazon Web Services (AWS), Microsoft Azure, and Nectar. These cloud environments offer scalable computing power and storage capabilities, high reliability and availability, and advanced services for data analytics and machine learning. The cloud layer supports containerization for consistent deployment of ReinFog components and typically handles computationally intensive tasks, large-scale data processing, and long-term data storage.

\subsubsection{Edge/Fog Layer}
The Edge/Fog Layer serves as an intermediate computing tier between the cloud and IoT devices. This layer comprises various computing devices, including laptops, desktops with varying computational capabilities, and single-board computers (e.g., Raspberry Pi). These devices are strategically positioned closer to the data source, enabling reduced latency for time-sensitive applications, local data processing and filtering, and improved privacy and security by keeping sensitive data local. The Edge/Fog layer connects to the Cloud layer via Internet connections, allowing for seamless data transfer and task offloading when needed. It also supports containerization for flexible deployment of ReinFog software components, playing a crucial role in facilitating real-time applications and reducing the computational burden on both the cloud and IoT devices.

\subsubsection{IoT Layer}
The IoT Layer comprises the various end devices and sensors that collect data and interact with the physical environment. This layer includes smart home devices (e.g., thermostats, security cameras), wearable devices, industrial sensors and actuators, connected vehicles, and environmental monitoring sensors. These devices are characterized by limited computational resources and power constraints, with direct interaction with the physical world through sensing and actuation. For connectivity, these IoT devices employ various short-range communication protocols (e.g., WiFi, Bluetooth, Zigbee, 5G) to connect with the Edge/Fog layer, while Internet-based connections facilitate communication with the Cloud layer. The IoT layer is the primary source of data in the ReinFog ecosystem, driving the need for efficient resource management and IoT application scheduling. We assume each IoT application in this layer can consist of one or multiple interdependent IoT tasks that need to be efficiently scheduled.

\subsection{Software Architecture}
\begin{figure}[pos=t]
\includegraphics[width=\linewidth]{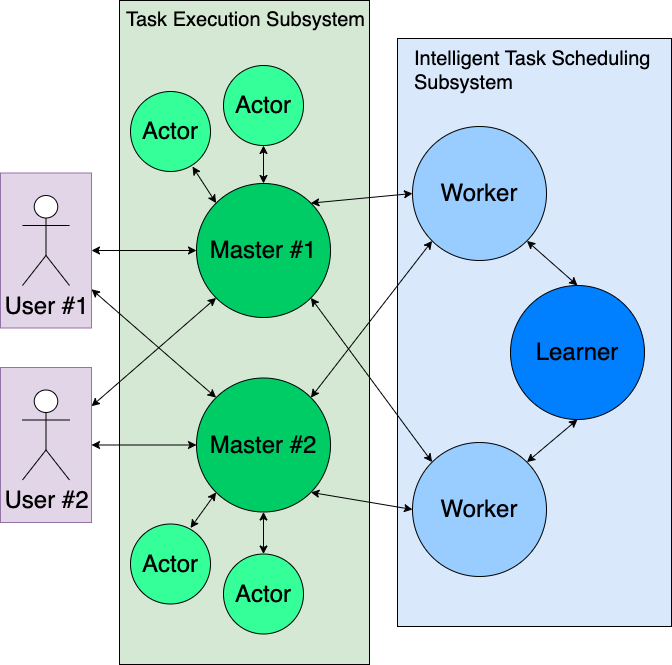}
\caption{High-level software architecture of ReinFog} \label{design}
\end{figure}

The software architecture of ReinFog is designed to efficiently manage and schedule IoT tasks in heterogeneous edge/fog and cloud environments. As illustrated in Figure \ref{design}, the framework consists of two primary subsystems, along with user interfaces for IoT application submission.

\subsubsection{Task Execution Subsystem}
This subsystem forms the operational core of ReinFog, which is responsible for managing the execution of IoT applications across the distributed environment. It comprises:

\begin{itemize}
    \item \textbf{Master:} Acting as the central coordinator, the Master manages overall system operations. It receives IoT application requests, analyzes task dependencies, and constructs directed acyclic graphs (DAGs) to represent application structures. The Master interacts with the Intelligent Task Scheduling Subsystem for optimal scheduling decisions, considering both task dependencies and resource availability. It oversees execution, ensuring correct task order. The architecture supports multiple Master instances for scalability and fault tolerance.
    
    \item \textbf{Actor:} Distributed across various nodes in the system, Actors are responsible for executing assigned tasks and managing local resources. They implement the actual task execution, monitor local system performance, and handle data transfer between tasks when necessary. Actors continuously report resource utilization, task progress, and completion status back to their associated Master. This real-time feedback enables dynamic resource management and adaptive IoT task scheduling, allowing the system to respond efficiently to changing workloads and environmental conditions.
\end{itemize}

The Task Execution Subsystem handles critical functions such as application analysis, task distribution, execution monitoring, and result aggregation, ensuring efficient utilization of available resources across the heterogeneous computing environment.

\subsubsection{Intelligent Task Scheduling Subsystem}
This subsystem represents ReinFog's core innovation, leveraging DRL to make intelligent IoT scheduling decisions. It consists of:

\begin{itemize}
    \item \textbf{Worker:} Workers gather system state information from the Task Execution Subsystem and generate IoT task scheduling decisions based on the learned policy. They communicate these decisions back to the Task Execution Subsystem for implementation. Workers continuously refine their decision-making process, adapting to the system's dynamic nature through distributed learning.
    
    \item \textbf{Learner:} Centrally located, the Learner aggregates experiences or gradients from Workers and optimizes the global DRL technique. The Learner periodically distributes the refined policy to all Workers, ensuring system-wide improvement in decision-making. This continuous optimization enables ReinFog to adapt to changing conditions and enhance overall IoT task scheduling efficiency.
\end{itemize}

The Intelligent Task Scheduling Subsystem continuously learns from the system's performance, adapting to changing conditions over time. This adaptive approach allows ReinFog to efficiently handle the dynamic nature of IoT workloads and the heterogeneity of computing resources.

\subsubsection{ReinFog Operation Workflow}
In the ReinFog framework, the workflow begins when users submit IoT applications through dedicated interfaces. These applications typically comprise multiple interdependent tasks that require coordinated execution. Upon receiving user submissions, the Master in the Task Execution Subsystem serves as the first point of contact, processing the applications by analyzing task dependencies and constructing DAGs to represent the execution structure. The Master then forwards scheduling requests to the Worker in the Intelligent Task Scheduling Subsystem, along with the system states and task characteristics. Based on this real-time information, the Worker generates IoT task scheduling decisions using DRL policies and sends these decisions back to the Master. Upon receiving these decisions, the Master coordinates its managed Actors to execute the specific tasks, with Actors handling the actual task execution and providing status feedback. Throughout this process, the Learner in the Intelligent Task Scheduling Subsystem works in parallel to optimize the DRL policies globally based on execution results and system states, continuously distributing improved policies to all Workers to enhance the system's scheduling efficiency and adaptability.

This architecture allows ReinFog to balance the load across available resources and optimize overall system performance. By integrating DRL-based decision-making with traditional task execution mechanisms, ReinFog can effectively adapt to the complex and dynamic nature of modern IoT, edge/fog and cloud computing environments, providing a robust and efficient solution for resource management and IoT application scheduling.

\section{ReinFog Design and Implementation}
To achieve adaptive resource management in edge/fog and cloud environments, we implement the ReinFog framework by extending the core components of the FogBus2 framework and implementing and integrating the new Intelligent Task Scheduling Subsystem. We chose FogBus2 as the foundation for ReinFog because it already implements many of the basic functionalities required in our Task Execution Subsystem.

FogBus2 is a lightweight and distributed container-based framework for integrating heterogeneous IoT systems with edge/fog and cloud environments. It comprises five main components that align well with our Task Execution Subsystem requirements:

\begin{itemize}
    \item \textbf{User:} handles environmental data collection and actuation control, similar to our user interface for IoT application submission.
    \item \textbf{Master:} manages IoT applications and scheduling, which forms the basis of our Master component in the Task Execution Subsystem.
    \item \textbf{Actor:} performs host resource profiling, aligning with our Actor component's responsibilities.
    \item \textbf{Task Executor:} executes submitted IoT applications, fitting directly into our execution model.
    \item \textbf{Remote Logger:} provides persistent log storage, supporting our system's monitoring and analysis needs.
\end{itemize}

\begin{figure*}[pos=t]
\includegraphics[width=\textwidth]{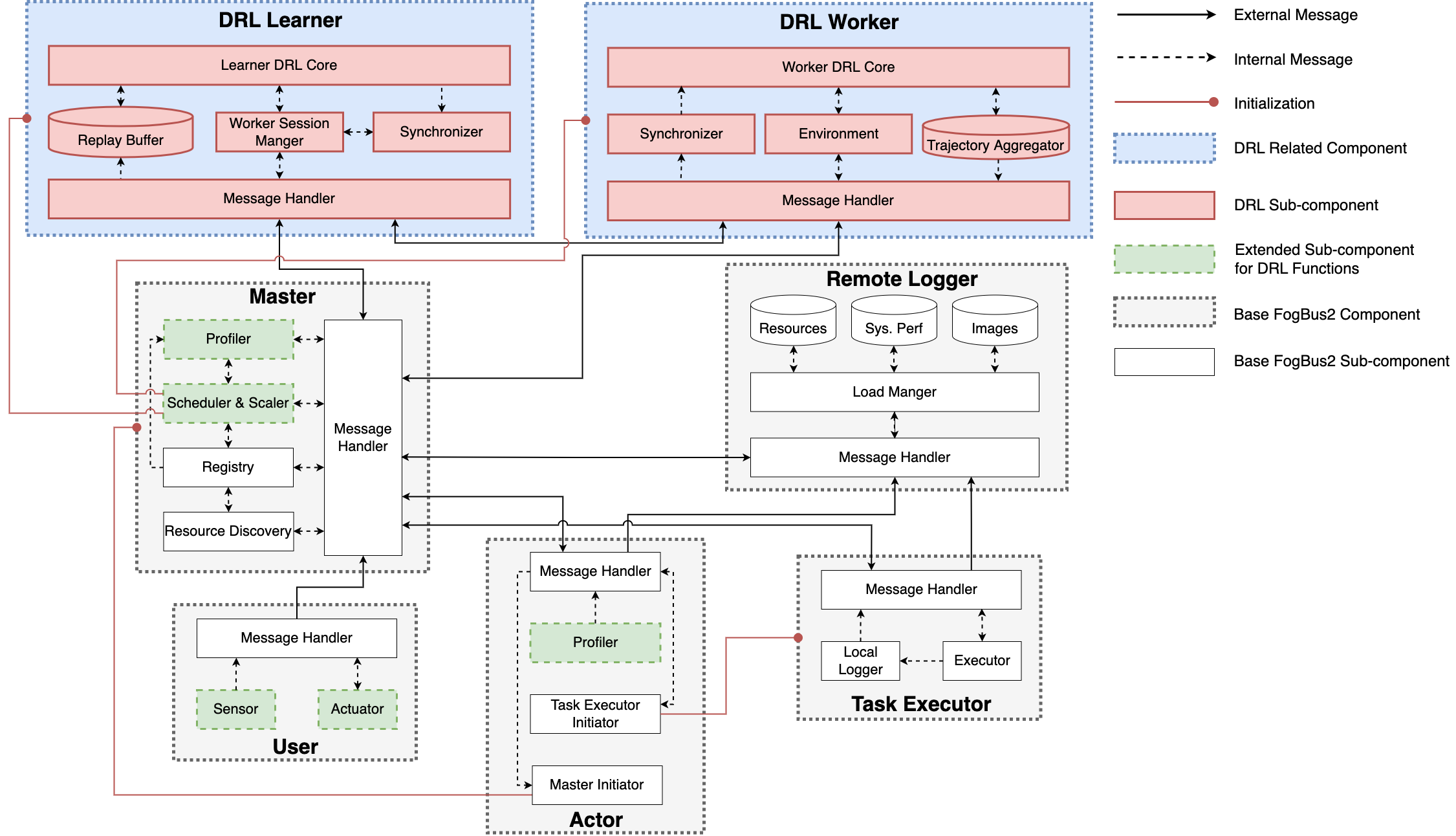}
\caption{ReinFog design overview} \label{overview}
\end{figure*}

By leveraging FogBus2, we are able to focus our efforts on developing our DRL-based scheduling subsystem to handle complex, dependency-aware IoT applications. This approach allowed us to build upon a proven foundation while integrating our novel Intelligent Task Scheduling Subsystem with its DRL capabilities. The overall design of ReinFog, showing both the extended FogBus2 components and our new DRL components, is illustrated in Fig.~\ref{overview}.

\subsection{ReinFog DRL Components}
\begin{figure*}[pos=t]
\includegraphics[width=\textwidth]{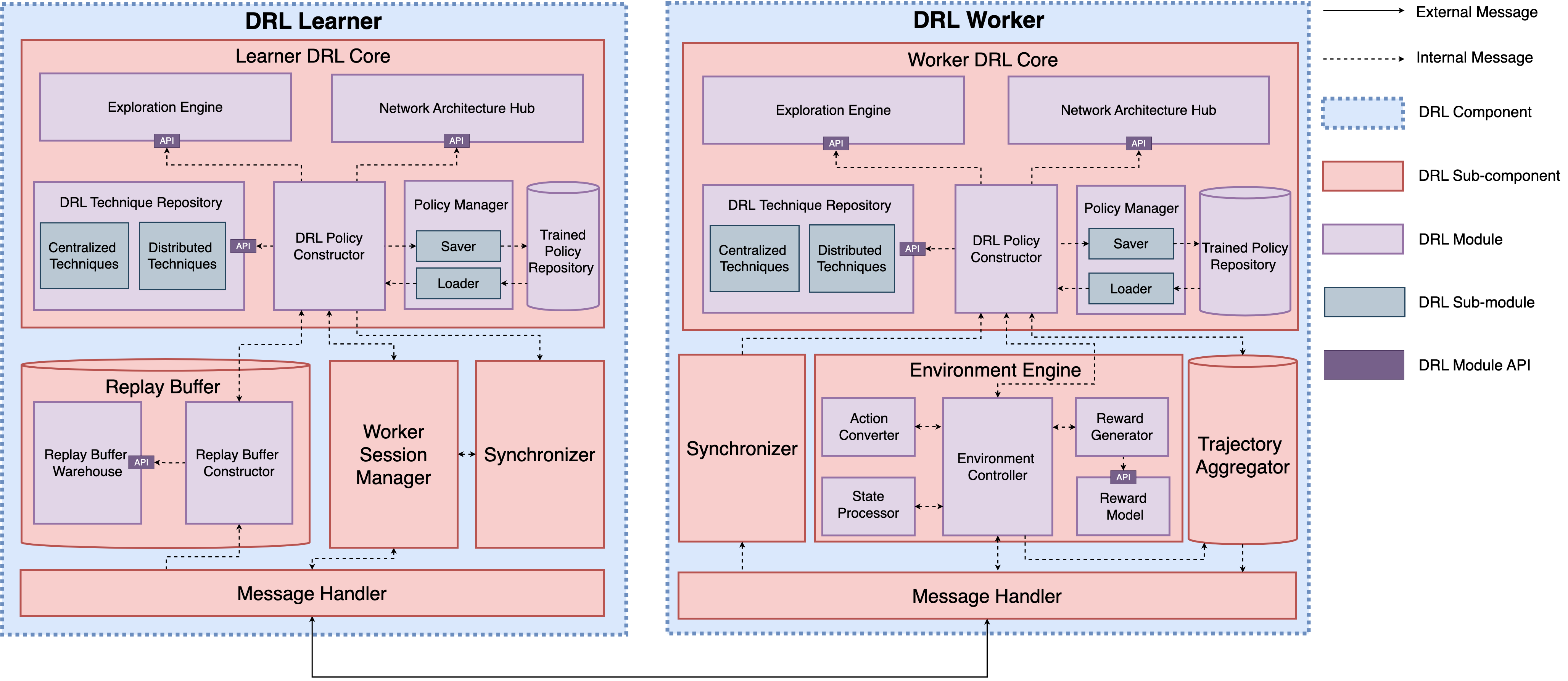}
\caption{Distributed DRL components design} \label{ddrlsys}
\end{figure*}

In this section, we present the design and organization of our DRL components: the DRL Learner and the DRL Worker. Each component is modularly designed and encompasses several sub-components and modules, which interact through well-defined APIs and internal messages. The overall design is depicted in Fig.~\ref{ddrlsys}.
\par
\subsubsection{DRL Learner}
The DRL Learner component is responsible for training the DRL techniques and managing the learning process. It comprises the following sub-components:

\paragraph{\textbf{Learner DRL Core:}}
The Learner DRL Core is central to policy learning and optimization. It includes the following modules:

\begin{enumerate}

\item \textbf{Exploration Engine:} This module implements and manages exploration strategies during the training process to balance exploration and exploitation. It incorporates established methods such as $\epsilon$-greedy and Ornstein–Uhlenbeck process noise, while also providing an extensible mechanism for integrating additional exploration algorithms. The Exploration Engine exposes a well-defined API, allowing other components to interact with it efficiently. Through this API, it dynamically interacts with the Network Architecture Hub and the DRL Technique Repository via the DRL Policy Constructor, enabling it to access and utilize current technique states and network architectures. This API-driven design enhances modularity and facilitates easy integration of new exploration strategies. 

\item \textbf{Network Architecture Hub:} This module manages and maintains base neural networks for various DRL techniques. It supports a wide range of architectures including:
\begin{itemize}
\item Deep Neural Networks (DNNs)
\item Recurrent Neural Networks (RNNs)
\item Long Short-Term Memory Networks (LSTMs)
\item Transformers
\end{itemize}
These diverse architectures enable the hub to accommodate different types of tasks. The module exposes APIs for querying and updating network architectures, facilitating seamless integration and dynamic adaptability. Designed with extensibility in mind, it allows for easy incorporation of new neural network architectures and policy optimization methods in the future.

\item \textbf{DRL Technique Repository:} This module efficiently manages and maintains various DRL techniques. It is divided into Centralized Techniques and Distributed Techniques sub-modules, accommodating different user requirements and computational architectures. The repository provides two integration mechanisms. The native integration mechanism enables direct implementation of DRL techniques within the framework, offering optimal performance and full customization flexibility. ReinFog already includes native implementations of several representative DRL techniques, such as DQN, PPO, A3C, and IMPALA. In parallel, the library-based integration mechanism enables seamless incorporation of external DRL libraries through standardized interfaces. ReinFog has already integrated techniques including SAC and R2D2 via the Ray library. This dual-mechanism design allows developers to either build custom DRL techniques within the framework or directly leverage existing implementations, ensuring flexibility, maintainability, and performance consistency across heterogeneous computing environments.

\item \textbf{DRL Policy Constructor:} This module is central to building and optimizing DRL policies. It orchestrates interactions with multiple modules through well-defined APIs to create and refine effective DRL strategies. The Constructor leverages the Network Architecture Hub's API for base network construction and accesses the DRL Technique Repository via API for policy development. It interacts with the Exploration Engine through its API to balance exploration and exploitation during the learning process. For policy persistence, the Constructor integrates with the Policy Manager, enabling saving and loading of policies. To enhance learning efficiency, it collaborates with the Replay Buffer for experience sampling. The DRL Policy Constructor also coordinates parallel learning processes via the Worker Session Manager and ensures policy synchronization through the Synchronizer.  

\item \textbf{Policy Manager:} This module is responsible for comprehensive DRL policy administration. It consists of two key sub-modules: the Saver and the Loader. The Saver sub-module ensures version-controlled, persistent storage of optimized policies in the Trained Policy Repository. The Loader sub-module, on the other hand, retrieves policies from the Repository as needed for further refinement or evaluation. The Policy Manager interfaces directly with the DRL Policy Constructor, enabling seamless integration of policy persistence operations within the overall learning process. 

\item \textbf{Trained Policy Repository:} This module functions as a centralized storage for optimized and validated DRL policies. It interfaces directly with the Policy Manager's Saver and Loader sub-modules, enabling efficient storage and retrieval of policies. The Repository maintains a structured archive of policies, supporting version control and rapid access for DRL experiments. It enhances the overall efficiency and effectiveness of policy management in the ReinFog framework.
\end{enumerate}

\paragraph{\textbf{Replay Buffer:}}
The Replay Buffer enhances learning efficiency and stability by storing and reusing experiences. It consists of two key modules:
\begin{enumerate}
\item \textbf{Replay Buffer Warehouse:} This module stores agent-environment interaction experiences. It features both random and reservoir sampling buffers. Random sampling helps break temporal correlations in the data, while reservoir sampling maintains a fixed-size buffer suitable for streaming or unbounded data. The Warehouse also supports flexible integration of additional sampling algorithms.
\item \textbf{Replay Buffer Constructor:} This module manages the creation, maintenance, and updating of buffers. It interfaces with the Replay Buffer Warehouse through its API to facilitate the storage, retrieval, and sampling of experiences. It also interacts directly with the DRL Policy Constructor, supporting efficient experience utilization in the learning process. 
\end{enumerate}

\paragraph{\textbf{Worker Session Manager:}}
The Worker Session Manager is responsible for managing communication across multiple DRL Workers in a distributed environment. It handles thread creation, maintenance, and DRL Worker communication, ensuring fast and reliable transmission of data and instructions. This sub-component plays a crucial role in coordinating parallel learning processes and maintaining efficient distributed operations within the ReinFog framework.

\paragraph{\textbf{Synchronizer:}}
The Synchronizer is tasked with synchronizing training across the distributed learning environment. It coordinates policy updates, gradients, and parameters between the DRL Learner and DRL Workers, maintaining a unified learning environment. Through close collaboration with the Worker Session Manager, the Synchronizer ensures consistency in the learning process, facilitating effective distributed learning in ReinFog.

\paragraph{\textbf{Message Handler:}}
The Message Handler manages inter-component communications within the ReinFog framework. It receives and processes incoming messages, efficiently routing them to the appropriate internal sub-components. It serves as a central communication hub, facilitating effective information exchange between various components of the ReinFog framework. By ensuring smooth and organized message flow, the Message Handler maintains system coherence and optimizes overall operational efficiency.

\subsubsection{DRL Worker}
The DRL Worker component is responsible for interacting with the environment, collecting and processing data to support policy learning and optimization within the ReinFog framework. It shares core functionalities with the DRL Learner, with the Worker DRL Core mirroring the architecture of the Learner DRL Core to ensure consistent functionality and coordination. Despite these similarities, the DRL Worker incorporates several unique sub-components: an Environment Engine for direct interaction with the learning environment, a Trajectory Aggregator for efficient collection and processing of experience data, and a Synchronizer adapted for DRL Worker-specific synchronization tasks, functioning differently from the Synchronizer in the DRL Learner. In centralized learning scenarios, the DRL Worker can assume the role of the DRL Learner, enabling direct policy updates for DRL techniques. This comprehensive design enables flexible deployment of the DRL Worker in both centralized and distributed learning contexts, enhancing the overall adaptability and efficiency of the ReinFog framework.

\paragraph{\textbf{Environment Engine:}}
The Environment Engine manages interactions between the learning components and the environment through several key modules:
\begin{enumerate}
\item \textbf{State Processor:} This module transforms raw environmental data into a format suitable for decision-making and learning processes.
\item \textbf{Action Converter:} This module translates generated actions into environment-specific commands, ensuring proper execution of decisions.
\item \textbf{Reward Generator:} This module processes reward parameters and utilizes the Reward Model's API to calculate rewards.
\item \textbf{Reward Model:} This module defines reward calculation methods, maintains consistency in reward generation, and supports the extension of additional reward functions.
\item \textbf{Environment Controller:} This module oversees the overall interaction process, managing the flow of actions and states, and coordinating the integration of other modules.
\end{enumerate}
This modular design ensures a smooth and continuous interaction cycle, facilitating efficient learning and adaptation within the ReinFog framework.

\paragraph{\textbf{Trajectory Aggregator:}}
The Trajectory Aggregator is responsible for collecting and processing trajectories generated from interactions with the environment. These trajectories, consisting of sequences of states, actions, and rewards, encapsulate experiential data over time. The aggregator maintains a well-structured and accessible repository of these experiences for training and evaluation purposes. In specific techniques such as A3C, where gradients are calculated within DRL Workers, the Trajectory Aggregator collects and transmits these gradients to the DRL Learner for updating the global policy. Moreover, in centralized learning scenarios where policies are optimized within the DRL Worker component, the Trajectory Aggregator can adapt to function as a Replay Buffer, storing and sampling trajectory data to support policy training and updates. This versatile design enables the Trajectory Aggregator to support various learning paradigms and techniques within the ReinFog framework, enhancing flexibility and efficiency in different operational contexts.

\paragraph{\textbf{Synchronizer:}}
The Synchronizer in the DRL Worker maintains consistency between local and global policies in distributed learning scenarios. It periodically obtains the latest policy parameters from the DRL Learner and updates the local policies accordingly. This mechanism facilitates efficient knowledge sharing, allowing DRL Workers to make decisions based on up-to-date global knowledge while contributing to the overall learning process. The Synchronizer is essential for balancing local exploration and global exploitation within the ReinFog framework.

\subsection{Extended FogBus2 Sub-components}
To enable FogBus2 to work seamlessly with DRL capabilities, we have extended several of its sub-components. This section introduces these extended sub-components, as illustrated in Fig.~\ref{overview}.

\subsubsection{Extended Scheduler \& Scaler}
\begin{figure*}[pos=t]
\includegraphics[width=\textwidth]{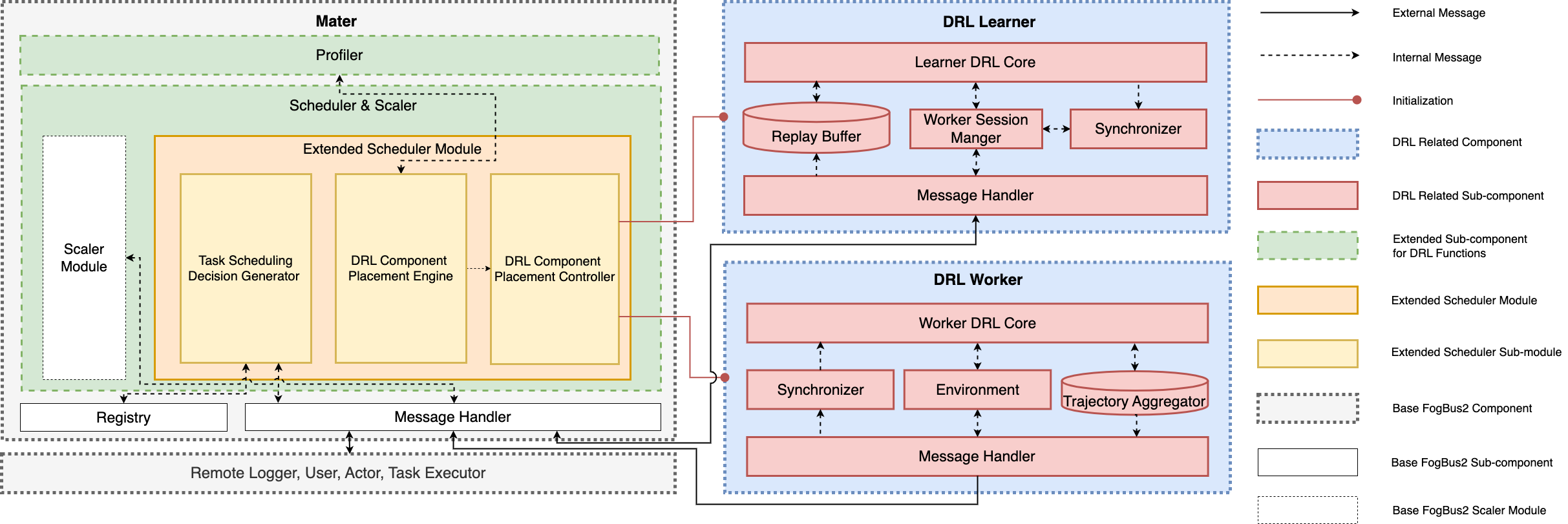}
\caption{Extended FogBus2 scheduler module} \label{extension}
\end{figure*}

The Scheduler \& Scaler sub-component has been extended to incorporate DRL capabilities, with the primary focus on enhancing the Scheduler Module. This extension enables coordination with the newly introduced DRL components. Fig.~\ref{extension} illustrates the detailed structure of the Extended Scheduler Module within the Scheduler \& Scaler sub-component.

\paragraph{\textbf{DRL Component Placement Engine:}}
The DRL Component Placement Engine generates strategies for placing DRL Learners and DRL Workers. It incorporates several algorithms:
\begin{itemize}
\item \textbf{GA}: Inspired by natural selection, GA evolves a population of potential solutions over generations.
\item \textbf{FA}: Based on the flashing behavior of fireflies, FA uses attraction and movement towards brighter solutions.
\item \textbf{PSO}: Mimicking the social behavior of bird flocking, PSO updates solutions based on personal and global best experiences.
\item \textbf{MADCP}: Our proposed algorithm that combines GA, FA, and PSO to leverage their respective strengths for more efficient DRL component placement.
\end{itemize}
In addition, the Engine supports the integration of additional algorithms to address diverse DRL component placement requirements and optimization objectives.

\paragraph{\textbf{DRL Component Placement Controller:}}
The DRL Component Placement Controller retrieves placement strategies from the DRL Component Placement Engine and applies them to place DRL Learners and DRL Workers. This sub-module ensures that the DRL components are properly set up according to the generated strategies, facilitating the efficient start of the learning process.

\paragraph{\textbf{Task Scheduling Decision Generator:}}
The Task Scheduling Decision Generator communicates with DRL components to obtain IoT task scheduling decisions. It then manages the deployment of these tasks based on the received decisions. This sub-module bridges the gap between DRL components and practical task execution within the ReinFog framework.

\subsubsection{Extended Profiler}
The Profiler has been extended in both the Master and Actor components to provide more comprehensive system monitoring. In the Master, the extended Profiler now collects and analyzes system-wide performance metrics, providing a holistic view of the entire distributed environment. This includes monitoring resource utilization patterns, network latencies, and overall system throughput. In the Actor, the Profiler has been augmented to gather more detailed node-specific information, such as CPU and memory usage, task execution times, and local network conditions. These enhancements enable more accurate modeling of the system state, which is crucial for the DRL components to make informed IoT task scheduling decisions.

\subsubsection{Extended User's Sensor and Actuator}
The User component's Sensor and Actuator have been extended to enhance data collection and performance monitoring capabilities, supporting the integration of DRL in ReinFog. These extensions focus on gathering more detailed information about IoT device operations and application performance. Key enhancements include the ability to monitor detailed energy consumption patterns of IoT devices, fine-grained operational timing such as task execution and completion times, as well as device-specific characteristics like processing capabilities and storage capacity. Additionally, the extensions provide real-time data on network conditions and connectivity status of IoT devices.

\subsection{Centralized and Distributed Deployment}
ReinFog supports both centralized and distributed deployment for DRL techniques. This section details how ReinFog components are utilized in these two deployment patterns, with specific examples to illustrate the implementation of each deployment.

\subsubsection{Centralized DRL Techniques Deployment}
Centralized DRL techniques deployment is essential in ReinFog for scenarios where the environment complexity is moderate and the system can benefit from simplified architecture and reduced communication overhead. In centralized deployment, a single DRL Worker handles both learning and decision-making processes, suitable for scenarios with relatively stable workload patterns or when system resources are limited.

\paragraph{\textbf{Centralized DRL Deployment Scheme in ReinFog:}}
In centralized deployment, the learning process is conducted directly within the DRL Worker, eliminating the need for a separate DRL Learner. The Environment Engine collects environmental data from the ReinFog Master and processes them for training.  The Trajectory Aggregator will assume
the role of the Replay Buffer, which stores and samples experiences for policy optimization. The Worker DRL Core contains all the necessary modules for both learning and decision-making. This design centralizes all environmental data collection and policy updating within the DRL Worker, creating a self-contained unit that operates independently.

\paragraph{\textbf{Sample Illustration:}}
To demonstrate the centralized deployment, we utilize DQN \cite{mnih2015human} as an example. DQN is a widely applied DRL technique that combines Q-learning with DNNs, employing a primary network for action selection and a separate target network for value estimation to stabilize training and improve learning efficiency in sequential decision-making problems \cite{arulkumaran2017deep}. In our implementation, the DRL Worker's Environment Engine receives environmental data from the ReinFog Master as a scheduling request and processes them into suitable formats for DQN utilization. The Worker DRL Core maintains two neural networks: the primary network for action selection and the target network for stable Q-value estimation. Actions are selected based on the current Q-values and sent to the Master for IoT task scheduling. The scheduling results and rewards are processed by the Environment Engine and stored in the Trajectory Aggregator. The Worker DRL Core then samples mini-batches of experiences directly from the Trajectory Aggregator and performs Q-learning updates. 

\subsubsection{Distributed DRL Techniques Deployment}
Distributed DRL techniques deployment is essential in ReinFog for addressing complex, large-scale environments where centralized approaches may be inefficient. This deployment pattern can harness the computational capabilities of heterogeneous edge/fog and cloud resources through multiple DRL Workers collaborating with a single or multiple DRL Learners, enabling parallel training and distributed learning to deal with dynamic workloads that demand rapid adaptation.

\paragraph{\textbf{Distributed DRL Deployment Scheme in ReinFog:}}
In distributed deployment, ReinFog employs multiple DRL Workers collaborating with a single or multiple DRL Learners to enable parallel training and decision-making. Each DRL Worker operates independently, with its Environment Engine collecting and processing environmental data from the Master and its DRL Core generating local decisions based on the current policy. The Trajectory Aggregator in each DRL Worker collects and processes local experiences for transmission. The Worker Session Manager coordinates the communication flow, enabling DRL Workers to transmit their processed experiences to the central DRL Learner. The DRL Learner stores these experiences in its Replay Buffer. The Learner DRL Core then performs global policy optimization using experiences stored and sampled from the Replay Buffer. After policy updates, the Synchronizer ensures policy consistency by distributing the latest global policy to all DRL Workers. This distributed architecture enables scalable learning across multiple nodes while maintaining policy coherence.

\paragraph{\textbf{Sample Illustration:}}
To illustrate the distributed deployment, we present IMPALA as an example. IMPALA is a highly scalable distributed DRL technique that efficiently handles large-scale learning by employing multiple actors running in parallel with a centralized learner, using importance sampling and V-trace off-policy correction to maintain stability in the learning process. In our implementation of IMPALA, multiple DRL Workers operate in parallel, each with its Environment Engine processing environmental data received from the ReinFog Master. The Worker DRL Core maintains local copies of the actor and critic networks for policy representation. Based on these local networks, the Worker DRL Core generates actions for IoT task scheduling. The scheduling results and rewards are processed locally by the Environment Engine, and each DRL Worker's Trajectory Aggregator collects these experiences. The collected experiences are then transmitted to the DRL Learner through worker sessions managed by the Worker Session Manager, and stored in the Replay Buffer. The Learner DRL Core maintains the global actor and critic networks for policy optimization. It samples experiences from the Replay Buffer, computes importance sampling ratios to address the discrepancy between behavior and target policies, and employs V-trace for off-policy correction to update these global networks. The updated global policy parameters are then distributed to all DRL Workers through the Synchronizer, ensuring consistency across the distributed system.

\subsection{Native and Library-based Integration}
ReinFog supports two primary mechanisms for integrating DRL techniques: native integration and library-based integration. The native integration mechanism enables direct implementation of DRL techniques within the framework, offering optimal performance and customization capabilities. The library-based integration mechanism allows seamless incorporation of external DRL libraries, providing access to established implementations while maintaining framework compatibility. This dual integration approach ensures that ReinFog can accommodate diverse research needs while leveraging existing DRL solutions. In this section, we detail both integration mechanisms and provide examples of their implementation within ReinFog.

\subsubsection{Native DRL Techniques Integration}
The native integrationmechanism is essential to ReinFog as it offers substantial advantages in performance, maintainability, and flexibility. Directly implementing DRL techniques within the framework can optimize performance, particularly in dynamic edge/fog and cloud environments where real-time responsiveness is crucial. This mechanism also minimizes dependencies on external libraries, enhancing stability and compatibility across various deployment scenarios. Furthermore, native integration provides researchers with the flexibility to implement and integrate their own custom DRL techniques, allowing for tailored solutions to specific research problems. 

\paragraph{\textbf{Native Integration Mechanism:}}
To support native integration of DRL techniques, ReinFog provides a comprehensive framework through well-defined interfaces and modular design, to offer maximum flexibility while maintaining consistency and efficiency. The Network Architecture Hub is responsible for neural network architecture definition and management. It maintains a collection of base network architectures (e.g., DNNs, RNNs, LSTMs, Transformers) and allows researchers to define custom architectures through a consistent interface. Building upon these network architectures, the DRL Technique Repository is responsible for managing DRL techniques, providing an extensible interface through a base class. This base class defines essential abstract methods including policy network initialization, action selection, loss computation, and parameter updates. These abstractions enable researchers to implement new DRL techniques. ReinFog's Exploration Engine supports native integration through a pluggable exploration strategy interface, offering built-in configurability for common strategies (e.g., $\epsilon$-greedy and Ornstein–Uhlenbeck process noise) while enabling custom strategy implementation. The Replay Buffer enables custom buffer implementation through a standard interface. It provides several built-in buffer types (e.g., random sampling and reservoir sampling) while allowing researchers to define flexible sampling strategies and integrate specialized buffer types.

\paragraph{\textbf{Sample Illustration:}}
To demonstrate the native integration mechanism, we describe how to implement a customized version of IMPALA that leverages ReinFog's advanced features. The Network Architecture Hub is utilized to construct two Transformer-based neural networks. The actor network processes the input states and outputs action probabilities for IoT task scheduling decisions, while the critic network evaluates these states to guide the learning process. IMPALA's core algorithm is implemented by extending the base class in the DRL Technique Repository, where the V-trace off-policy correction and importance sampling calculations are defined. The Exploration Engine is configured to use Ornstein-Uhlenbeck noise for action exploration. The Replay Buffer is set up with reservoir sampling to efficiently store and manage experiences. Through these configurations and implementations, IMPALA operates as a fully functional distributed DRL technique within the ReinFog framework.

\subsubsection{Library-based DRL Techniques Integration}
The library-based integration mechanism serves as another crucial feature of ReinFog, offering development efficiency, algorithmic diversity, and research flexibility. By importing well-established DRL libraries, researchers can leverage proven algorithms without reimplementation, significantly reducing development time while ensuring reliable performance. This mechanism also enriches ReinFog with diverse DRL techniques, enabling it to handle various scheduling scenarios in heterogeneous edge/fog and cloud environments. Moreover, it empowers researchers to integrate and experiment with different DRL libraries based on their specific research requirements.

\paragraph{\textbf{Library-based Integration Mechanism:}}
To support external DRL libraries integration, ReinFog provides a standardized interface through the DRL Technique Repository. This interface defines essential methods for bridging external libraries with ReinFog's environment. Specifically, the interface allows researchers to implement state preprocessing to convert ReinFog's system state representations into formats compatible with external libraries, action translation to map library-generated actions back to ReinFog's scheduling decisions, and reward signal adaptation to ensure proper learning feedback. The DRL Technique Repository manages these transformations, enabling external DRL techniques to operate seamlessly within ReinFog's resource management framework while maintaining their original implementations.

\paragraph{\textbf{Sample Illustration:}}
To demonstrate the library-based integration mechanism, we provide an example of integrating Recurrent Replay Distributed DQN (R2D2) \cite{kapturowski2018recurrent} from the Ray\footnote{https://www.ray.io/} library. Ray library is chosen for its high-performance distributed computing features and comprehensive DRL techniques suite. R2D2 is a distributed DRL technique that extends DQN by incorporating RNNs and a replay system to handle partial observability and temporal dependencies in sequential decision-making problems. Through the interface provided in the DRL Technique Repository, we implement state preprocessing to convert ReinFog's scheduling environment states (e.g., node resources, IoT task characteristics, and network conditions) into R2D2's required tensor format. The action translation mechanism transforms R2D2’s output probabilities into concrete scheduling decisions within ReinFog, effectively guiding the scheduling of IoT tasks across available nodes. For reward signal adaptation, the implementation processes ReinFog's performance metrics (e.g., scheduling result, response time, energy consumption) into a scalar reward value suitable for R2D2's learning process. These implementations enable R2D2 to effectively learn and make scheduling decisions within ReinFog while preserving its original recurrent replay-based learning mechanism.

\section{MADCP: A Memetic Algorithm for DRL Component Placement}
In ReinFog, effective DRL-based IoT application scheduling requires careful placement of various components (e.g., DRL Learners and DRL Workers) across different nodes in the heterogeneous computing environment. Poor DRL component placement decisions can lead to increased communication overhead, inefficient resource utilization, and degraded learning performance. To solve this challenge, ReinFog offers a mechanism for DRL component placement. While traditional meta-heuristic algorithms could be applied to this placement problem, they show specific limitations. GA provides robust exploration capabilities but may converge slowly in complex solution spaces \cite{renders1996hybrid}. FA excels at local search refinement but can be trapped in local optima \cite{wu2020improved}. PSO offers efficient global search but may lack fine-tuning abilities in local regions \cite{moradi2016hybrid}. To address this critical placement challenge while overcoming these algorithmic limitations, we propose MADCP, a Memetic Algorithm for DRL Component Placement that combines the strengths of GA, FA, and PSO. In ReinFog, MADCP is integrated into the DRL Component Placement Engine of the extended Scheduler module (see Fig. \ref{extension}). It is invoked before DRL training to determine efficient placement of DRL Learners and Workers across nodes.

In this section, we first define the optimization problem we are addressing along with the MADCP. We also explain how our proposed MADCP combines the strengths of three methods: GA, FA, and PSO. In addition, we present a comprehensive analysis of the computational complexity of MADCP, examining its initialization and iterative optimization phases.

\subsection{Optimization Problem Definition}
The MADCP is designed to address a complex optimization problem in distributed computing environments. The primary goal is to efficiently place DRL components across heterogeneous computing nodes to optimize overall system performance.

\subsubsection{Problem Formulation}
Consider a set of DRL components $\mathcal{C} = \{C_1, C_2, \ldots, C_m\}$ that need to be assigned to a set of heterogeneous computing nodes $\mathcal{N} = \{N_1, N_2, \dots, N_n\}$. Each node has different computational capabilities, memory sizes, and energy consumption rates.

Each component $C_i$ has a computational requirement $U_i$ (e.g., CPU cycles), a memory requirement $M_i$, and a deadline or time constraint $D_i$. Similarly, each node $N_j$ is characterized by a computational capacity $P_j$, and available memory $A_j$.

\subsubsection{Objective Function}
The objective is to find an optimal assignment of DRL components to computing nodes that minimizes the total operation time and energy consumption while meeting all time and resource constraints. We define the objective function as:
\begin{align}
\text{Minimize} \quad F = & \sum_{i=1}^{m} \sum_{j=1}^{n} x_{ij} \big( \omega_1 \cdot O(C_i, N_j) \notag \\
& + \omega_2 \cdot E(C_i, N_j) \big),
\end{align}

Here, $x_{ij}$ is a binary variable indicating whether component $C_i$ is assigned to node $N_j$:
\begin{equation}
x_{ij} =
\begin{cases}
1, & \text{if component } C_i \text{ is assigned to node } N_j, \\
0, & \text{otherwise}.
\end{cases}
\end{equation}
$O(C_i, N_j)$ is the operation time of component $C_i$ on node $N_j$, $E(C_i, N_j)$ is the energy consumed by node $N_j$ during the operation time $O(C_i, N_j)$, and $\omega_1$ and $\omega_2$ are weighting factors balancing the importance of operation time and energy consumption.

\subsubsection{Constraints}
The optimization problem is subject to the following constraints:

\paragraph{\textbf{Resource Constraints:}}
For each node $N_j$, the total computational and memory requirements of the assigned components should not exceed its capacity:
\begin{equation}
\sum_{i=1}^{m} x_{ij} U_i \leq P_j, \quad \forall N_j \in \mathcal{N},
\end{equation}

\begin{equation}
\sum_{i=1}^{m} x_{ij} M_i \leq A_j, \quad \forall N_j \in \mathcal{N}.
\end{equation}

\paragraph{\textbf{Deadline Constraints:}}
Each component must complete the operation within its deadline:
\begin{align}
O(C_i, N_j) \leq D_i, &\quad \forall C_i \in \mathcal{C}, \notag \\
&\quad \forall N_j \in \mathcal{N} \text{ where } x_{ij} = 1.
\end{align}

\paragraph{\textbf{Assignment Constraints:}}
Each component is assigned to exactly one node:
\begin{align}
x_{ij} \in \{0, 1\}, \quad &\forall C_i \in \mathcal{C}, \quad \forall N_j \in \mathcal{N}, \\
\sum_{j=1}^{n} x_{ij} = 1, \quad &\forall C_i \in \mathcal{C}.
\end{align}

\subsubsection{Mathematical Model}
Combining the objective function and the constraints, the optimization problem can be formulated as:
\begin{align}
\text{Minimize} \quad & F = \sum_{i=1}^{m} \sum_{j=1}^{n} x_{ij} \Big( \omega_1 \cdot O(C_i, N_j) \notag \\
& \quad\quad\quad\quad + \omega_2 \cdot E(C_i, N_j) \Big) \\
\text{Subject to} \quad & \sum_{i=1}^{m} x_{ij} U_i \leq P_j, \quad \forall N_j \in \mathcal{N} \\
& \sum_{i=1}^{m} x_{ij} M_i \leq A_j, \quad \forall N_j \in \mathcal{N}
\end{align}

\begin{align}
& O(C_i, N_j) \leq D_i, \quad \forall C_i \in \mathcal{C}, \notag \\
& \qquad\quad \forall N_j \in \mathcal{N} \text{ where } x_{ij} = 1 \\
& x_{ij} \in \{0, 1\}, \quad \forall C_i \in \mathcal{C}, \quad \forall N_j \in \mathcal{N} \\
& \sum_{j=1}^{n} x_{ij} = 1, \quad \forall C_i \in \mathcal{C}
\end{align}

This problem belongs to a class of combinatorial optimization challenges, where the objective is to efficiently allocate resources under multiple constraints, making it computationally intensive (NP-hard). The complexity arises from the exponential number of possible component-to-node assignments as the number of components and nodes increases. Traditional exact algorithms become impractical for large-scale instances because they require prohibitive computational time to explore all possible solutions. Moreover, single-method heuristic algorithms often fail to adequately balance exploration and exploitation in the vast solution space, leading to suboptimal results.

\subsection{MADCP}
Given the limitations of existing methods in addressing large-scale, NP-hard optimization problems, we propose a novel memetic algorithm named MADCP to find near-optimal solutions within reasonable computational time for DRL components (e.g., DRL Learners and DRL Workers) placement. MADCP synergistically combines the strengths of three optimization algorithms:

\begin{enumerate}
\item \textbf{GA:} Provides robust exploration capabilities by simulating the process of natural selection, allowing the algorithm to broadly traverse the solution space and maintain genetic diversity among solutions.

\item \textbf{FA:} Enhances local search by modeling the flashing behavior of fireflies, enabling fine-tuning in promising areas and improving exploitation of the solution space.

\item \textbf{PSO:} Contributes efficient global optimization and rapid convergence by simulating social behavior patterns of organisms, allowing particles (solutions) to adjust their positions based on personal and global best experiences.
\end{enumerate}

By integrating these algorithms, MADCP achieves a balanced search strategy that mitigates the shortcomings of individual techniques, leading to improved performance in solving complex optimization problems. Algorithm~\ref{alg:MADCP} provides a detailed description of MADCP.
\begin{algorithm}[!t]
\scriptsize
\caption{MADCP}
\label{alg:MADCP}
\SetAlgoLined
\KwIn{\\
$compParameters$ -- parameters defining the components\\
$nodeParameters$ -- parameters defining the nodes\\
$populationSize$ -- number of individuals in the population\\
$generations$ -- number of iterations\\
$numOperations$ -- number of genetic operations per generation\\
$mutationRate$ -- probability of mutation\\
$FAParameters$ -- parameters $(\alpha, \beta, \gamma)$ for FA\\
$PSOParameters$ -- parameters $(w, c_1, c_2)$ for PSO
}
\KwOut{Optimal assignment of components to nodes}

\tcc{Initialization}
$components \leftarrow$ \textbf{CreateComponents}($compParameters$)\;
$nodes \leftarrow$ \textbf{CreateNodes}($nodeParameters$)\;
$population \leftarrow$ \textbf{GeneratePopulation}($populationSize$, $components$, $nodes$)\;
$velocities \leftarrow$ \textbf{InitializeVelocities}($populationSize$, $components$)\;
$personalOpt \leftarrow population$\;
$globalOpt \leftarrow \arg\max_{p \in population}$ \textbf{CalculateFitness}($p$, $components$)\;

\For{$i \leftarrow 1$ \KwTo $generations$}{
    \tcc{Fitness Evaluation}
    \ForEach{$p \in population$}{
        $fitnessValues[p] \leftarrow$ \textbf{CalculateFitness}($p$, $components$)\;
    }
    $newPopulation \leftarrow \emptyset$\;
    
    \tcc{Genetic Algorithm Operations}
    \For{$j \leftarrow 1$ \KwTo $numOperations$}{
        \tcc{Selection}
        $parent1, parent2 \leftarrow$ \textbf{SelectParents}($population$, $fitnessValues$)\;
        \tcc{Crossover}
        $child1, child2 \leftarrow$ \textbf{Crossover}($parent1$, $parent2$)\;
        \tcc{Mutation}
        $child1 \leftarrow$ \textbf{Mutate}($child1$, $mutationRate$, $nodes$)\;
        $child2 \leftarrow$ \textbf{Mutate}($child2$, $mutationRate$, $nodes$)\;
        \textbf{Add} $child1$ \textbf{and} $child2$ \textbf{to} $newPopulation$\;
    }
    $population \leftarrow newPopulation$\;
    
    \tcc{Firefly Algorithm Movement}
    $population \leftarrow$ \textbf{FireflyMovement}($population$, $components$, $\alpha$, $\beta$, $\gamma$)\;
    
    \tcc{Particle Swarm Optimization Update}
    $population$, $velocities \leftarrow$ \textbf{PSOUpdate}($population$, $velocities$, $personalOpt$, $globalOpt$, $w$, $c_1$, $c_2$)\;
    
    \tcc{Update Personal and Global Best Solutions}
    \ForEach{$p \in population$}{
        \If{\textbf{CalculateFitness}($p$, $components$) $>$ \textbf{CalculateFitness}($personalOpt[p]$, $components$)}{
            $personalOpt[p] \leftarrow p$\;
        }
    }
    $globalOpt \leftarrow$ \textbf{UpdateGlobalOpt}($population$, $globalOpt$, $components$)\;
}
\Return $globalOpt$\;
\end{algorithm}

In the initialization phase (Lines 1--6), the algorithm begins by generating the set of components and nodes based on the provided parameters. The function \textbf{CreateComponents} initializes the components $\mathcal{C}$, each with specific computational and memory requirements, as well as deadlines. Similarly, \textbf{CreateNodes} initializes the nodes $\mathcal{N}$, each characterized by their computational capacities and available memory. An initial population of candidate solutions is generated using \textbf{GeneratePopulation}, where each individual represents a potential assignment of components to nodes. For the PSO phase, velocities for each individual are initialized using \textbf{InitializeVelocities}. The personal best solutions $personalOpt$ are initially set to the current population, and the global best solution $globalOpt$ is determined by evaluating the fitness of each individual using \textbf{CalculateFitness} and selecting the one with the highest fitness value.

The algorithm then enters the main loop (Lines 7--28), iterating over a predefined number of generations to evolve the population towards optimality. At the beginning of each generation, the fitness of each individual in the population is calculated using \textbf{CalculateFitness} (Lines 8--10). The fitness function is designed based on the objective function defined in the optimization problem, considering execution time and energy consumption.

In the Genetic Algorithm operations (Lines 12--18), selection, crossover, and mutation are performed to generate a new population. Pairs of parent individuals are selected from the current population based on their fitness values using \textbf{SelectParents} (Line 13). The selected parents undergo crossover using \textbf{Crossover} (Line 14) to produce offspring, combining parts of the parents' solutions to create new individuals and promote exploration of the solution space. The offspring are then subjected to mutation using \textbf{Mutate} (Lines 15--16), introducing random changes to some genes (component assignments) with a probability defined by the mutation rate. This helps maintain genetic diversity and prevents premature convergence. The newly created children are added to form a new population (Line 19), replacing the old one.

After the Genetic Algorithm operations, the Firefly Algorithm is applied to the population using \textbf{FireflyMovement} (Line 20). Each individual moves towards better solutions based on attractiveness and randomization parameters, enhancing local search by exploiting promising regions in the solution space.

Subsequently, the Particle Swarm Optimization phase updates the positions and velocities of individuals using \textbf{PSOUpdate} (Line 21). Velocities are updated based on the inertia weight, cognitive coefficient, and social coefficient, guiding individuals towards their personal best and the global best solutions. Positions are then updated based on the new velocities, contributing to global optimization and convergence of the algorithm.

The personal best solutions $personalOpt$ are updated by comparing the current fitness of each individual with their personal best fitness (Lines 22--26); if an individual has achieved a better fitness, its personal best is updated accordingly. The global best solution $globalOpt$ is updated by identifying the best individual in the current population based on fitness evaluations (Line 27). After completing all generations, the algorithm returns the global best solution $globalOpt$ (Line 29), which represents the optimal assignment of components to nodes found by MADCP.

\subsection{Computational Complexity Analysis}
This section provides a detailed analysis of the computational complexity of the proposed MADCP. The complexity analysis is divided into two main parts: the initialization phase and the iterative optimization phase. By analyzing the time complexity of each phase, we derive the overall complexity of the algorithm.

\subsubsection{Initialization Phase}
The initialization phase consists of several operations, including the creation of components and nodes, population generation, velocity initialization, and initial fitness evaluation:

\begin{itemize}
    \item \textbf{Component and Node Creation}: The components and nodes are created based on the input parameters, where $M$ represents the number of components, and $N$ denotes the number of nodes. The complexity of this step is $O(M)$ for component creation and $O(N)$ for node creation, respectively.
    
    \item \textbf{Population Generation}: The initial population, with size $P$, is generated based on the number of components and nodes. Because each individual in the population represents a mapping between components and nodes, the complexity of this step is $O(P \times M)$.
    
    \item \textbf{Velocity Initialization}: The initial velocity of each individual is set based on the number of individuals in the population. The velocity initialization is performed by iterating over each component, so the complexity of this step is $O(P \times M)$. 
    
    \item \textbf{Fitness Evaluation}: The initial fitness of each individual in the population is calculated based on component assignments. The complexity for calculating the fitness of a single individual is $O(M)$. Therefore, evaluating the fitness of the entire population results in a complexity of $O(P \times M)$.
\end{itemize}

Combining the above components, the total complexity of the initialization phase can be expressed as:
\begin{equation}
O(M + N + P \times M).
\end{equation}

\subsubsection{Iterative Optimization Phase}
The iterative optimization phase consists of fitness evaluation, GA operations (selection, crossover, and mutation), FA-based movement, PSO update, and the update of personal and global best solutions.

\begin{itemize}
    \item \textbf{Fitness Evaluation}: In each generation, the fitness of every individual in the population is evaluated. The time complexity for evaluating the fitness of the entire population is $O(P \times M)$.
    
    \item \textbf{Genetic Algorithm Operations}: This step includes parent selection, crossover, and mutation:
    \begin{itemize}
        \item \textit{Selection}: Assuming a selection mechanism such as roulette wheel selection, the complexity is $O(P)$.
        \item \textit{Crossover}: The crossover operation generates two offspring, with a complexity of $O(M)$.
        \item \textit{Mutation}: Each offspring undergoes mutation with a complexity of $O(M)$.
    \end{itemize}
    Assume the GA operations are performed $P/2$ times per generation, the total complexity for GA operations is:
    \begin{equation}
    O\left(\frac{P}{2} \times (P + 2 \times M)\right) = O\left(P^2 + P \times M\right).
    \end{equation}
    
    \item \textbf{Firefly Algorithm Movement}: The FA-based movement operation involves comparing and updating the position of every individual relative to other individuals in the population. The complexity of this step is $O(P^2 \times M)$.
    
    \item \textbf{Particle Swarm Optimization Update}: The PSO update step adjusts the position and velocity of each individual in the population. The complexity of this step is $O(P \times M)$.
    
    \item \textbf{Update of Personal and Global Best Solutions}: This step involves comparing the fitness of each individual in the population with its historical best solution and updating the global best solution if necessary. The complexity for updating the personal and global best solutions is $O(P \times M)$.
\end{itemize}

The overall complexity of a single iteration is given by:
\begin{align}
O(P \times M) &+ O\left(P^2 + P \times M\right) + \nonumber \\
O(P^2 \times M) &+ O(P \times M) + O(P \times M),
\end{align}
which can be simplified to:
\begin{equation}
O(P^2 \times M).
\end{equation}
Assume the number of generations is $G$, the total complexity of the iterative optimization phase is:
\begin{equation}
O(G \times P^2 \times M).
\end{equation}

\subsubsection{Total Complexity}
By combining the complexity of the initialization phase and the iterative optimization phase, the overall time complexity of the MADCP is:
\begin{equation}
O(M + N + ( P \times M ) + ( G \times P^2 \times M)).
\end{equation}

In most practical scenarios, the complexity of the iterative optimization phase $O(G \times P^2 \times M)$ dominates the overall complexity, especially for larger values of $G$, $P$, or $M$. Therefore, the overall time complexity of the MADCP can be expressed as:
\begin{equation}
O(N + G \times P^2 \times M).
\end{equation}

The complexity analysis shows that although MADCP combines GA, FA, and PSO, the overall complexity does not significantly increase. Specifically, GA’s selection, crossover, and mutation operations have a complexity of $O(P^2 + P \times M)$, and PSO’s update operations have a complexity of $O(P \times M)$, both of which are relatively low and do not heavily impact the total complexity. The FA’s movement operation, with a complexity of $O(P^2 \times M)$, becomes the dominant factor as the population size increases. Instead, MADCP leverages GA to enhance population diversity, preventing premature convergence; FA to improve local search capabilities; and PSO to guide global convergence. This results in a superior optimization performance without adding substantial computational overhead. In essence, MADCP achieves a balanced integration of the strengths of each algorithm while maintaining manageable complexity, making it a highly effective optimization algorithm for complex optimization problems.

\section{Performance Evaluation}
This section presents a comprehensive evaluation of ReinFog's performance in addressing the complex IoT application scheduling problem across heterogeneous edge and cloud environments. In our performance evaluation, we adhered to established benchmarking practices to ensure the reliability of our results. We repeated each experiment 10 times per configuration to account for the variability in measurements. Moreover, we carefully controlled the experimental environment to minimize external disturbances and ensure consistent conditions for all tests. The final results presented are the averages of these repeated experiments, ensuring that they accurately reflect the typical system performance. 

\subsection{Experiments Setup}\label{exp}
We conduct our experiments in a heterogeneous computing environment comprising cloud, edge/fog, and IoT layers. The cloud layer consists of a multi-cloud setup including one AWS VM (Intel Xeon, 1 core @ 2.4GHz, 1GB RAM), one Azure VM (Intel Xeon, 1 core @ 2.3GHz, 1GB RAM), and 30 Nectar Cloud VMs (AMD EPYC; 16 VMs with 2 cores @ 2.0GHz, 8GB RAM; 10 VMs with 4 cores @ 2.0GHz, 16GB RAM; 4 VMs with 8 cores @ 2.0GHz, 32GB RAM). The edge/fog layer consists of an Apple Macbook Pro (macOS, Apple M1 Pro, 8 cores, 16GB RAM), a Dell laptop (Linux, Intel Core i7, 8 cores @ 2.3GHz, 16GB RAM), and a Raspberry Pi 3B (Raspberry Pi OS, Broadcom BCM2837, 4 cores @ 1.2GHz, 1GB RAM). Devices in the IoT layer are equipped with 2 cores @ 3.2GHz and 4GB RAM. The network configuration is characterized by the following latency/bandwidth metrics: IoT to Nectar 8-12ms/14-18MB/s, IoT to AWS 16-22ms/16-20MB/s, IoT to Azure 8-15ms/15-22MB/s, and IoT to edge/fog 1-4ms/130-150MB/s. We deploy four IoT applications in our experiments: Face Detection, Color Tracking, Face and Eye Detection, and Video Optical Character Recognition. These applications feature adjustable resolution and support both real-time and non-real-time processing.

Among the implemented/integrated DRL techniques, we use the following representative techniques to conduct our experiments. The configuration of each technique is discussed below:
\begin{itemize}
\item \textbf{IMPALA \cite{espeholt2018impala}}: IMPALA is a distributed DRL technique natively supported by ReinFog. We choose it for its exceptional scalability and efficiency in handling distributed learning scenarios \cite{espeholt2018impala}. In our experiments, it uses Ornstein–Uhlenbeck noise exploration strategy and the reservoir sampling replay buffer, and its network structure introduces Transformers. The DRL component placement algorithm is set to MADCP.
\item \textbf{A3C \cite{pmlr-v48-mniha16}}: A3C is a distributed DRL technique natively supported by ReinFog. We choose it for its asynchronous parallel learning architecture that enables efficient policy optimization through multiple independent actors operating concurrently \cite{pmlr-v48-mniha16}. In our experiments, it uses $\epsilon$-greedy exploration strategy, and its network structure introduces LSTMs. The DRL component placement algorithm is set to FA.
\item \textbf{PPO \cite{schulman2017proximal}}: PPO is a centralized DRL technique natively supported by ReinFog. We choose it for its stable policy optimization approach that uses a clipped objective function to prevent excessive policy updates while maintaining high sample efficiency \cite{tang2020discretizing}. In our experiments, it uses Ornstein–Uhlenbeck noise exploration strategy and the reservoir sampling replay buffer, and its network structure introduces LSTMs. The DRL component placement algorithm is set to PSO.
\item \textbf{DQN \cite{mnih2015human}}: DQN is a centralized DRL technique natively supported by ReinFog. We choose it as it is a fundamental and widely-adopted DRL technique that effectively combines Q-learning with DNNs through experience replay and target networks to stabilize training \cite{arulkumaran2017deep}. In our experiments, it uses $\epsilon$-greedy exploration strategy and the random sampling replay buffer, and its network structure introduces RNNs. The DRL component placement algorithm is set to GA. 
\item \textbf{R2D2 \cite{kapturowski2018recurrent}}: R2D2 is a distributed DRL technique imported by ReinFog from the Ray library. We choose it for its distributed architecture that extends DQN with RNNs to handle temporal dependencies while maintaining efficient parallel learning \cite{kapturowski2018recurrent}. In our experiments, the DRL component placement algorithm is set to MADCP.
\item \textbf{SAC \cite{haarnoja2018soft}}: SAC is a centralized DRL technique imported by ReinFog from the Ray library. We choose it for its maximum entropy approach that provides automatic tuning to balance exploration and exploitation, leading to robust and stable learning performance \cite{haarnoja2018soft}. In our experiments, the DRL component placement algorithm is set to PSO.
\item \textbf{OHNSGA \cite{deng2021fogbus2}}: OHNSGA is a GA-based meta-heuristic algorithm natively supported by FogBus2. We extended this technique to support multi-objective optimization for comparison with DRL techniques.
\end{itemize}

\subsection{Hyperparameters Settings}
 We conduct an extensive grid search to fine-tune the hyperparameters for each technique used in our experiments. The grid search involves systematically exploring various combinations of hyperparameters within predefined ranges to identify the optimal configuration for each technique.

For the neural network architecture, following the guidance from \cite{glorot2010understanding} and \cite{sun2019bert4rec}, we experiment with different numbers of fully connected layers (ranging from 2 to 5) and various combinations of hidden layer units ({16, 32, 64, 128, 256}). We find that a 3-layer architecture with [256, 256, 128] hidden units provided the best balance between model complexity and performance across most techniques.

The choice of activation function can significantly impact the learning dynamics. While ReLU (Rectified Linear Unit) is found to be effective for most techniques due to its non-linearity and reduced likelihood of vanishing gradients, we observe that TanH (Hyperbolic Tangent) works better for A3C, possibly due to its bounded output range.

Following the guidance from \cite{espeholt2018impala}, \cite{pmlr-v48-mniha16}, \cite{schulman2017proximal}, \cite{hessel2018rainbow}, and \cite{haarnoja2018soft}, learning rates are tuned within the range of 0.0001 to 0.01, with most techniques performing optimally around 0.001. The discount factor, which balances immediate and future rewards, is tuned within the range of 0.8 to 0.99. Most techniques show optimal performance with a discount factor of 0.99, while A3C performed better with a slightly lower value of 0.9.

For the OHNSGA algorithm, following the guidance from \cite{deb2002fast} and \cite{deb2013evolutionary}, which is not a DRL technique, we focus on tuning the population size and number of generations. The population size is varied from 50 to 500, and the number of generations is tested in the range of 50 to 200. After extensive experimentation, we find that a population size of 200 and 100 generations provides a good trade-off between solution quality and computational time. Table \ref{table:para} presents the optimal hyperparameters identified for each technique. 
\begin{table*}[pos=t]
\centering
\caption{Technique hyperparameters}
\label{table:para}
\resizebox{\textwidth}{!}{%
\begin{tabular}{|c|c|c|c|c|c|c|c|}
\hline
\textbf{Hyperparameters} & \textbf{IMPALA}   & \textbf{A3C}      & \textbf{PPO}      & \textbf{DQN}      & \textbf{R2D2}     & \textbf{SAC}      & \textbf{OHNSGA} \\ \hline
Fully Connected Layers   & 3                 & 3                 & 3                 & 3                 & 3                 & 3                 & -               \\ \hline
Hidden Layer Units       & {[}256,256,128{]} & {[}256,256,128{]} & {[}256,256,128{]} & {[}256,256,128{]} & {[}256,256,128{]} & {[}256,256,128{]} & -               \\ \hline
Activation Function      & ReLU              & TanH              & ReLU              & ReLU              & ReLU              & ReLU              & -               \\ \hline
Learning Rate            & 0.001             & 0.001             & 0.001             & 0.01              & 0.01              & 0.0001            & -               \\ \hline
Discount Factor          & 0.99              & 0.9               & 0.99              & 0.99              & 0.99              & 0.99              & -               \\ \hline
Population Size          & -                 & -                 & -                 & -                 & -                 & -                 & 200             \\ \hline
Generations Number       & -                 & -                 & -                 & -                 & -                 & -                 & 100             \\ \hline
\end{tabular}%
}
\end{table*}

In addition to the DRL techniques and OHNSGA, we also fine-tune the parameters for the DRL component placement algorithms: MADCP, GA, FA, and PSO. For MADCP, we tune the following parameters: population size (from 50 to 500), number of generations (from 50 to 200), crossover rate (from 0.6 to 1.0), light absorption coefficient ($\gamma$) (from 0.1 to 1.0), and inertia weight ($w$) (from 0.4 to 0.9). For GA, following the guidance from \cite{10.5555/645512.657403}, we tune the following parameters: population size (from 50 to 500), number of generations (from 50 to 200), crossover rate (from 0.6 to 1.0), mutation rate (from 0.01 to 0.05). For FA, following the guidance from \cite{yang2013firefly}, we focus on tuning: number of fireflies (from 20 to 200), randomization parameter ($\alpha$) (from 0.1 to 1.0), attractiveness coefficient ($\beta$) (from 0.1 to 1.0), and light absorption coefficient ($\gamma$) (from 0.1 to 1.0). For PSO, following the guidance from \cite{van2006study}, parameters are optimized as follows: swarm size (from 20 to 200), cognitive coefficient ($c_1$) (from 1.5 to 2.5), social coefficient ($c_2$) (from 1.5 to 2.5), and inertia weight ($w$) (from 0.4 to 0.9). 

After extensive experimentation, we identify the optimal parameter settings for each DRL component placement algorithm when paired with different DRL techniques. For MADCP, the optimal settings are population size: 200, number of generations: 100, crossover rate: 0.8, light absorption coefficient ($\gamma$): 0.5, and inertia weight ($w$): 0.7 when used with IMPALA. When paired with R2D2, most parameters remain the same, except for a slight change in inertia weight ($w$) to 0.8. For GA, which is used with DQN, the optimal parameters are population size: 200, number of generations: 100, crossover rate: 0.9, and mutation rate: 0.05. For FA, when used with A3C, the optimal settings are number of fireflies: 100, randomization parameter ($\alpha$): 0.2, attractiveness coefficient ($\beta$): 0.8, and light absorption coefficient ($\gamma$): 0.1. When used with PPO, only $\gamma$ changes to 0.2, while other parameters remain the same. PSO maintains the same parameters (swarm size: 100, cognitive coefficient ($c_1$): 2.0, social coefficient ($c_2$): 2.0, and inertia weight ($w$): 0.7) in both SAC and PPO. 

These fine-tuned parameters are used consistently across all experiments to ensure fair comparison. It's worth noting that while these hyperparameters yield the best overall performance in our experimental setup, the optimal configuration may vary depending on the specific characteristics of the deployment environment. 

\subsection{Evaluation Metrics}
In our experiments, we focus on three key metrics: Response Time ($\mathcal{RT}$), Energy Consumption ($\mathcal{EC}$), and Weighted Cost ($\mathcal{WC}$). These metrics are applied to a set of IoT applications $\mathcal{A}$, where each application $\mathcal{A}_i$ is modeled as a Directed Acyclic Graph (DAG) consisting of dependent tasks $\mathcal{T}^j_i$, scheduled across a set of nodes $\mathcal{N}$. $\mathcal{SC}$ denotes the set of scheduling configurations for these applications.

\paragraph{\textbf{Response Time ($\mathcal{RT}$):}}
This metric represents the total time required to process all applications. It is calculated as:
\begin{equation}
\mathcal{RT}(\mathcal{SC}) = \sum_{i=1}^{|\mathcal{A}|} \sum_{j=1}^{|\mathcal{A}_i|} \left( \mathcal{RT}(\mathcal{SC}^j_i) \times \mathcal{CP}(\mathcal{T}^j_i) \right),
\end{equation}
where $\mathcal{SC}^j_i$ refers to the scheduling configuration for task $\mathcal{T}^j_i$, and $\mathcal{CP}(\mathcal{T}^j_i)$ is a binary variable that indicates whether the task lies on the critical path of the application. The critical path determines the minimum time required for the application to complete.

\paragraph{\textbf{Energy Consumption ($\mathcal{EC}$):}}
This metric measures the total energy consumed during the execution of the applications:
\begin{equation}
\mathcal{EC}(\mathcal{SC}) = \sum_{i=1}^{|\mathcal{A}|} \sum_{j=1}^{|\mathcal{A}_i|} \mathcal{EC}(\mathcal{SC}^j_i),
\end{equation}
where $\mathcal{EC}(\mathcal{SC}^j_i)$ represents the energy consumed for task $\mathcal{T}^j_i$ under the given scheduling configuration $\mathcal{SC}^j_i$.

\paragraph{\textbf{Weighted Cost ($\mathcal{WC}$):}}
Weighted Cost ($\mathcal{WC}$) is a composite metric that balances both response time and energy consumption. It is defined as:
\begin{equation}
\mathcal{WC}(\mathcal{SC}) = w_1 \times Norm(\mathcal{RT}(\mathcal{SC})) + w_2 \times Norm(\mathcal{EC}(\mathcal{SC})),
\end{equation}
where $w_1$ and $w_2$ are weight parameters used to control the relative importance of Response Time ($\mathcal{RT}$) and Energy Consumption ($\mathcal{EC}$), respectively, with $w_1 + w_2 = 1$. The function $Norm(x)$ performs normalization of a value $x$ relative to a predefined baseline to eliminate the differences in units and scales between Response Time ($\mathcal{RT}$) and Energy Consumption ($\mathcal{EC}$). This ensures that both metrics contribute equally to the Weighted Cost ($\mathcal{WC}$) calculation, independent of their absolute values or units. In our experiments, we set $w_1 = 0.5$ and $w_2 = 0.5$, giving equal priority to both factors.

Based on the metrics, the reward function is defined as the negative Response Time ($\mathcal{RT}$), Energy Consumption ($\mathcal{EC}$), and Weighted Cost ($\mathcal{WC}$) if the application is successfully executed, or a penalty if it fails. The optimization goal is to minimize the three metrics, respectively, while satisfying the corresponding constraints related to application requirements and node processing capabilities.

\subsection{ReinFog Framework Performance Analysis}
In this section, we conduct a comprehensive evaluation of ReinFog's framework performance. Among the related works, we choose FogBus2 as our baseline framework for its most comprehensive generic capabilities for resource management framework and similar modular architecture. We first analyze the startup time of different components in ReinFog. We then evaluate their RAM usage, demonstrating the lightweight nature of our framework despite its DRL capabilities. Next, we assess the environmental impact by evaluating carbon dioxide (\(\text{CO}_2\)) emissions across the electricity generation patterns in different regions. Finally, we examine the scalability of the framework by analyzing its performance with varying numbers of DRL Workers.

\subsubsection{Framework Startup Time Analysis}
\begin{figure}[pos=t]
\includegraphics[width=\linewidth]{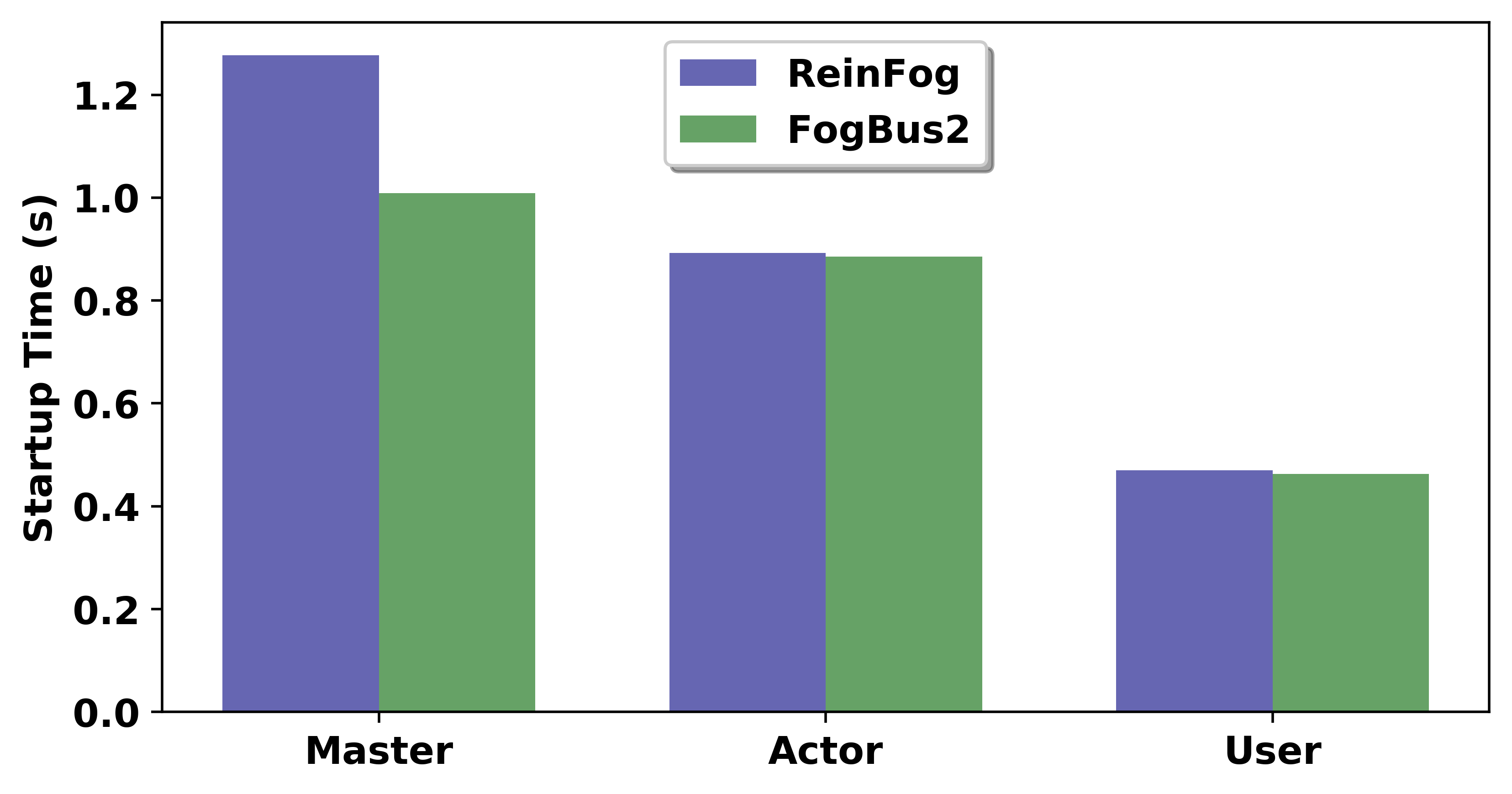}
\caption{Startup time comparison between ReinFog and FogBus2 components}\label{fig:stt}
\end{figure}

In this experiment, we compare the startup time of ReinFog and FogBus2 across three key components: Master, Actor, and User. As shown in Fig. \ref{fig:stt}, the startup time of the Actor (0.89s) and User (0.47s) components in ReinFog are nearly identical to those in FogBus2. However, the Master component in ReinFog requires approximately 0.25 seconds more startup time (1.26s vs 1.01s) compared to FogBus2's Master component, due to the initialization of DRL components. Despite this slight increase in the Master's startup time, the overall impact on system performance is minimal, demonstrating ReinFog's efficient design in integrating DRL capabilities while maintaining reasonable startup overhead.

\subsubsection{Framework RAM Usage Analysis}
\begin{figure}[pos=t]
\includegraphics[width=\linewidth]{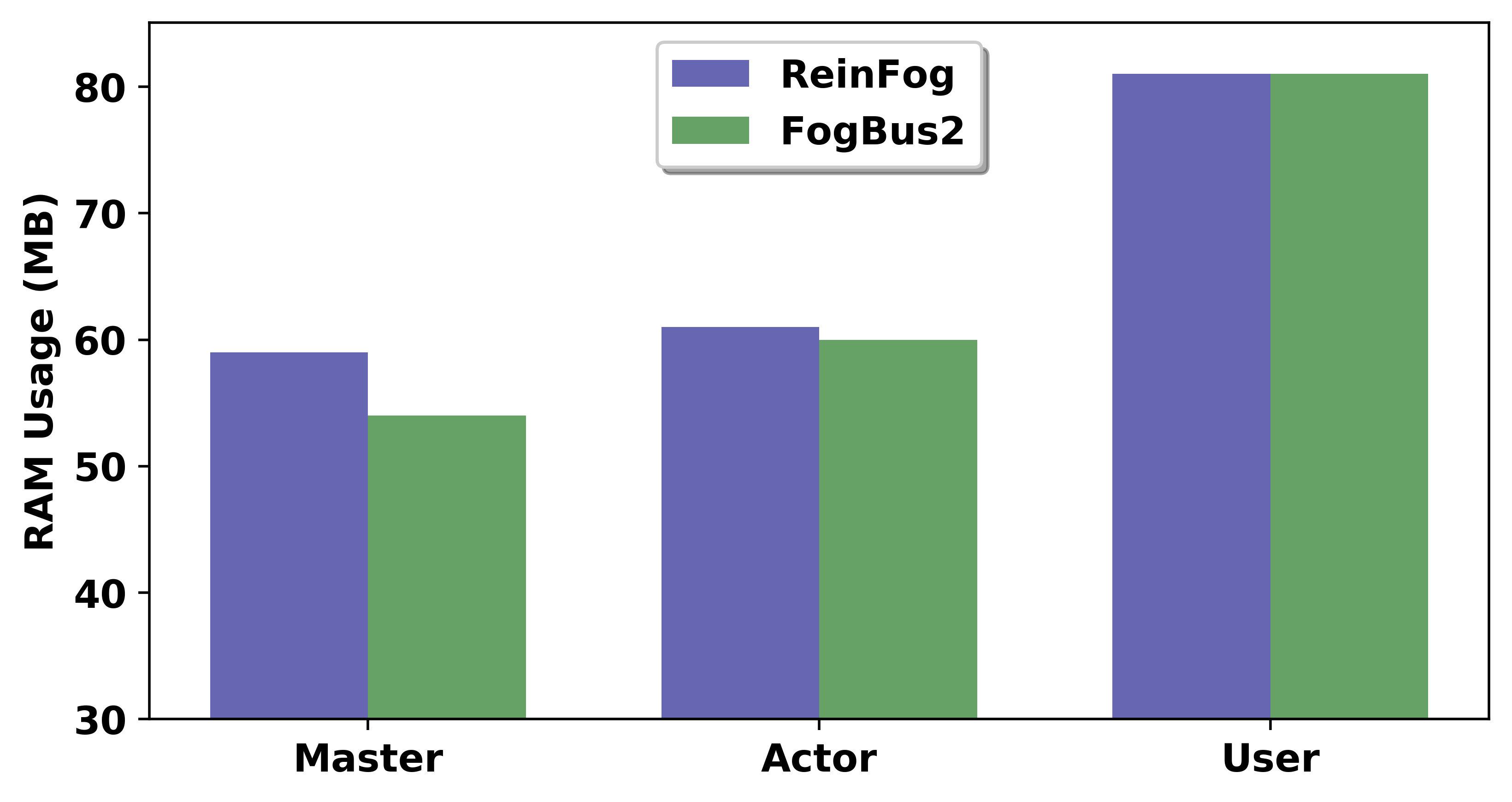}
\caption{RAM usage comparison between ReinFog and FogBus2 components}\label{fig:ram}
\end{figure}

In this experiment, we evaluate the RAM usage of ReinFog and FogBus2 across the Master, Actor, and User components. As illustrated in Fig. \ref{fig:ram}, the RAM usage of the User (81MB) and Actor (61MB) components exhibits minimal differences between ReinFog and FogBus2. The Master component in ReinFog utilizes approximately 5MB of additional memory compared to FogBus2 (59MB versus 54MB) due to the DRL components integration. This modest increase in RAM usage demonstrates that ReinFog maintains a lightweight design while incorporating advanced DRL capabilities.

\subsubsection{Framework Carbon Dioxide Emissions Analysis}\label{carbon}
\begin{figure}[pos=t]
\includegraphics[width=\linewidth]{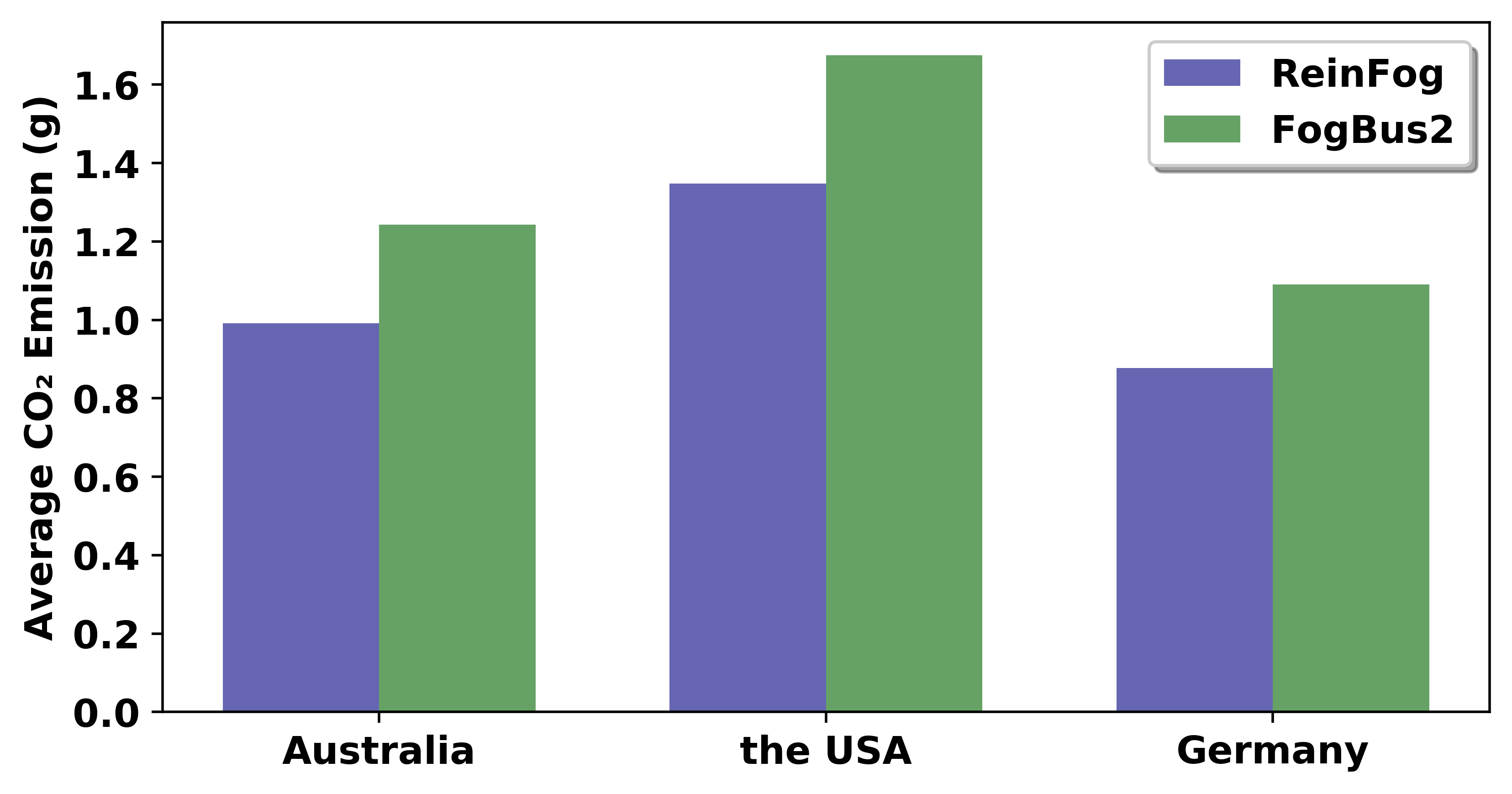}
\caption{Hourly CO2 emissions comparison between ReinFog and FogBus2 across different regions}\label{fig:aco}
\end{figure}

In this experiment, we evaluate the average \(\text{CO}_2\) emissions of ReinFog across all integrated DRL techniques and compare it with FogBus2 during one hour of IoT application processing. To enable a fair comparison, we extended FogBus2's scheduler with energy optimization capabilities. The primary objective is to assess whether the integration of DRL components in ReinFog introduces significant environmental overhead despite its energy optimization capabilities. The analysis considers the distinct electricity generation patterns of Australia, the United States of America (USA), and Germany to provide a comprehensive evaluation of environmental impact in regions with different energy source distributions. We record the electricity consumed and estimate the GHG emissions following the electricity generation patterns in Australia\footnote{https://www.energy.gov.au/data/electricity-generation}, the USA\footnote{https://www.eia.gov/tools/faqs}, and Germany\footnote{https://www.umweltbundesamt.de/themen/co2-emissionen-pro-kilowattstunde-strom-stiegen-in}, using the formula \(GHG = E \times \sum_{i}(e_i \times p_i)\), where \(E\) represents the total electricity consumed, \(e_i\) denotes the emission factor, and \(p_i\) indicates the proportion of the source $i$ (i.e., coal, gas) in producing electricity. As illustrated in Fig. \ref{fig:aco}, ReinFog consistently produces lower \(\text{CO}_2\) emissions than FogBus2 across all regions. Specifically, in Australia, ReinFog emits 0.99g compared to FogBus2's 1.24g. The difference is more pronounced in the USA, where ReinFog generates 1.35g versus FogBus2's 1.67g. Germany shows the lowest emissions for both frameworks (ReinFog: 0.88g, FogBus2: 1.09g), attributable to its higher proportion of renewable energy sources. These results demonstrate that ReinFog not only provides advanced DRL capabilities but also maintains a lower carbon footprint compared to FogBus2, with the actual environmental impact varying based on regional energy policies and power grid compositions.

\subsubsection{Framework Scalability Analysis}
\begin{figure}[pos=t]
\includegraphics[width=\linewidth]{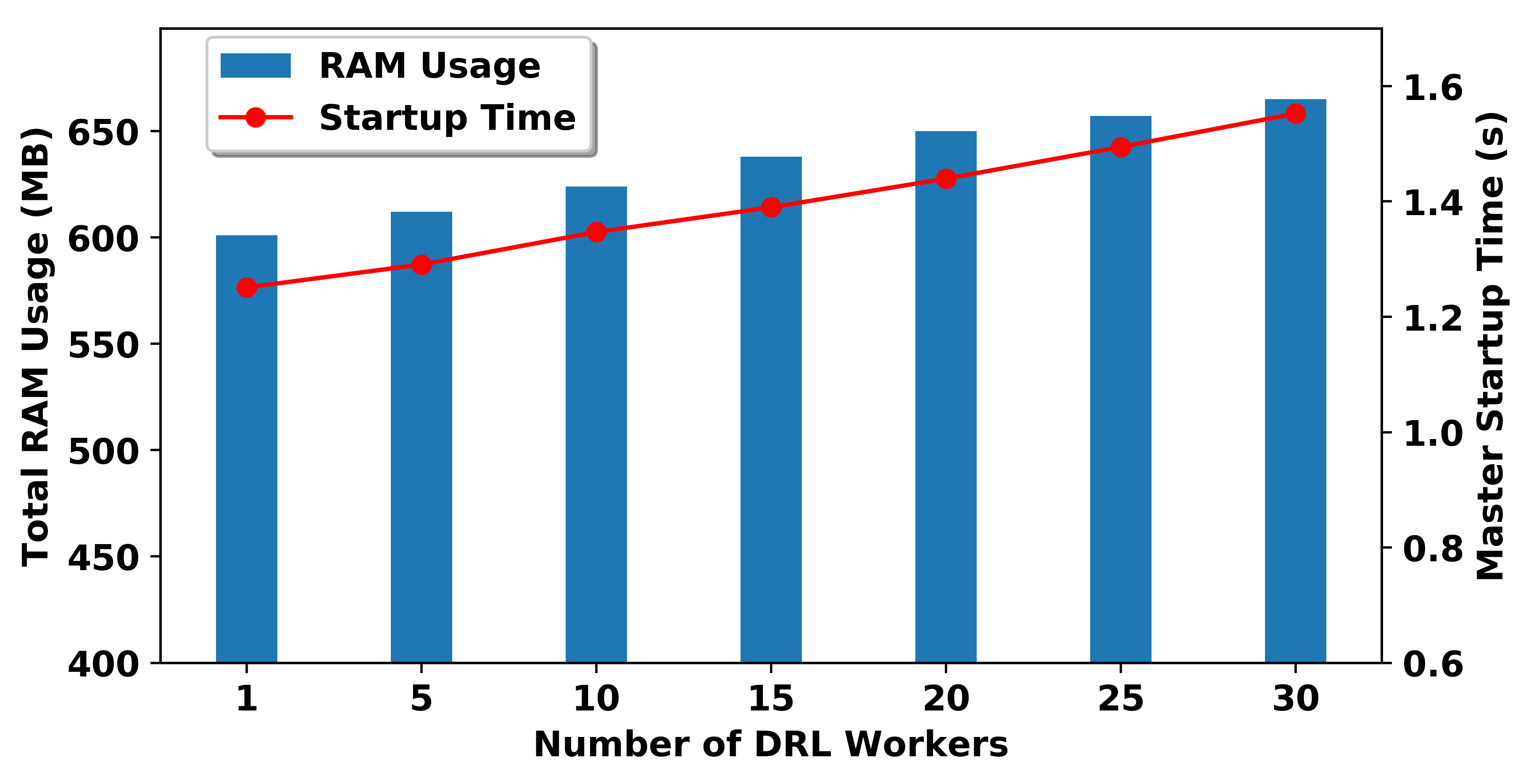}
\caption{Impact of increasing DRL Workers on ReinFog RAM usage and startup time}\label{fig:scr}
\end{figure}

In this experiment, we evaluate the scalability of ReinFog by systematically varying the number of DRL Workers from 1 to 30. As illustrated in Fig. \ref{fig:scr}, as the number of DRL Workers increases, both framework RAM usage and Master startup time show controlled growth. Specifically, the total framework RAM usage exhibits a linear increase from 600MB with one DRL Worker to 660MB with 30 DRL Workers, representing only a 10\% increase in memory consumption (i.e., roughly 2 MB of additional RAM usage per DRL Worker). Meanwhile, the Master startup time demonstrates a gradual increase from 1.25s to 1.56s, reflecting a modest rise of approximately 0.3s. These results indicate that ReinFog maintains efficient resource utilization and reasonable startup overhead even with a significant increase in the number of DRL Workers, demonstrating its excellent scalability for large-scale deployments.

\subsection{ReinFog DRL Scheduling Techniques Analysis}
In this section, we evaluate the performance of different DRL scheduling techniques integrated into ReinFog by comparing them with OHNSGA \cite{deng2021fogbus2}, the meta-heuristic scheduling approach originally proposed in FogBus2. This allows us to directly assess the benefits of DRL-based scheduling over a well-established non-learning baseline within a consistent system environment. We first analyze the convergence performance of various scheduling techniques during both training and evaluation phases across multiple metrics, including response time, energy consumption, and weighted cost. We then examine the computational overhead introduced by different scheduling techniques. Following this, we evaluate the scalability of these techniques by testing their performance across different numbers of nodes. Finally, we assess the environmental impact of different techniques by analyzing the \(\text{CO}_2\) emissions. 

\subsubsection{Scheduling Techniques Convergence Analysis}\label{sca}
\begin{figure*}[pos=htb]
\begin{subfigure}{0.33\textwidth}
  \centering
  \includegraphics[width=\linewidth]{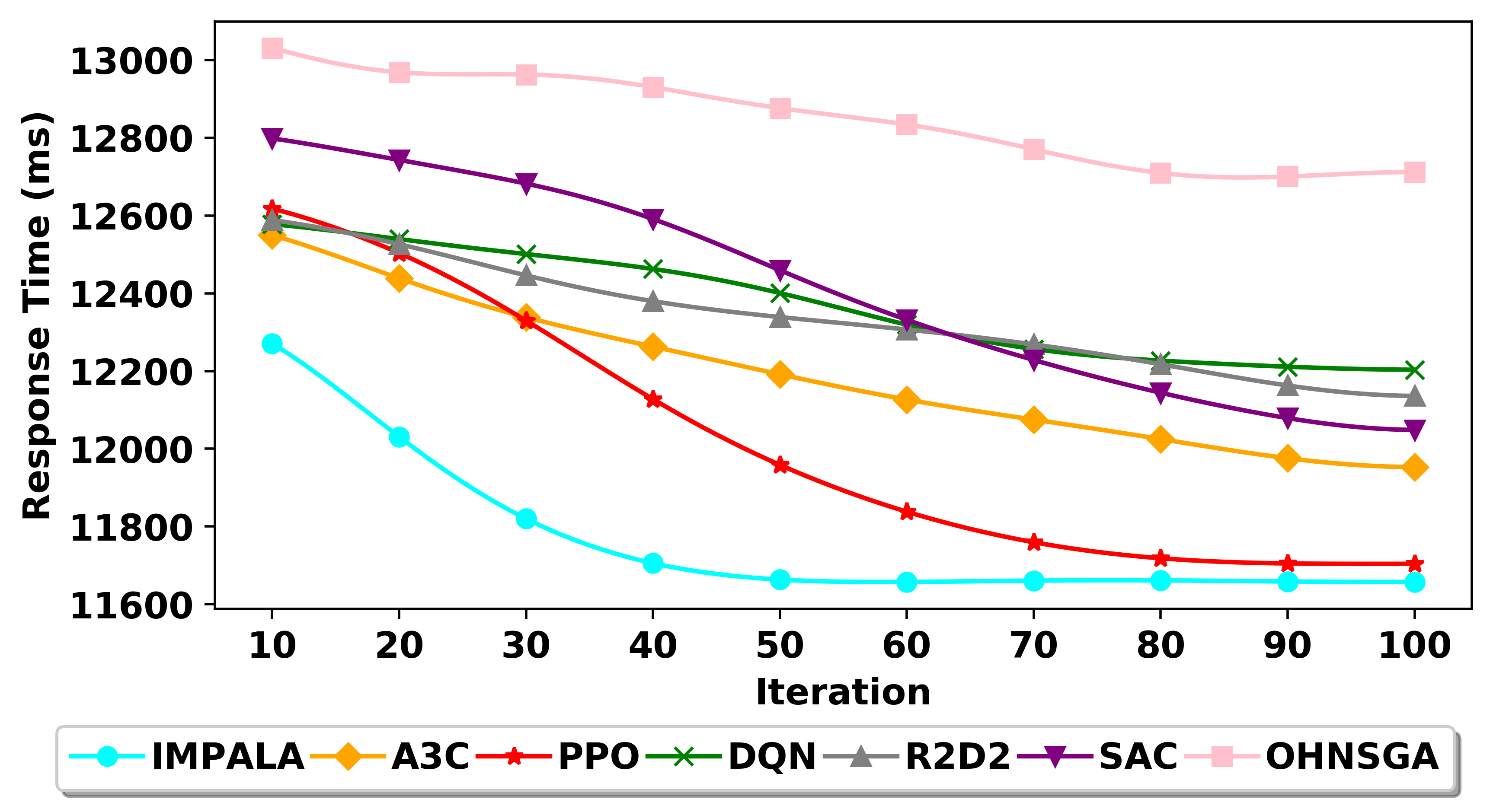}
  \caption{Response time}
  \label{fig:rtt}
\end{subfigure}%
\begin{subfigure}{0.33\textwidth}
  \centering
  \includegraphics[width=\linewidth]{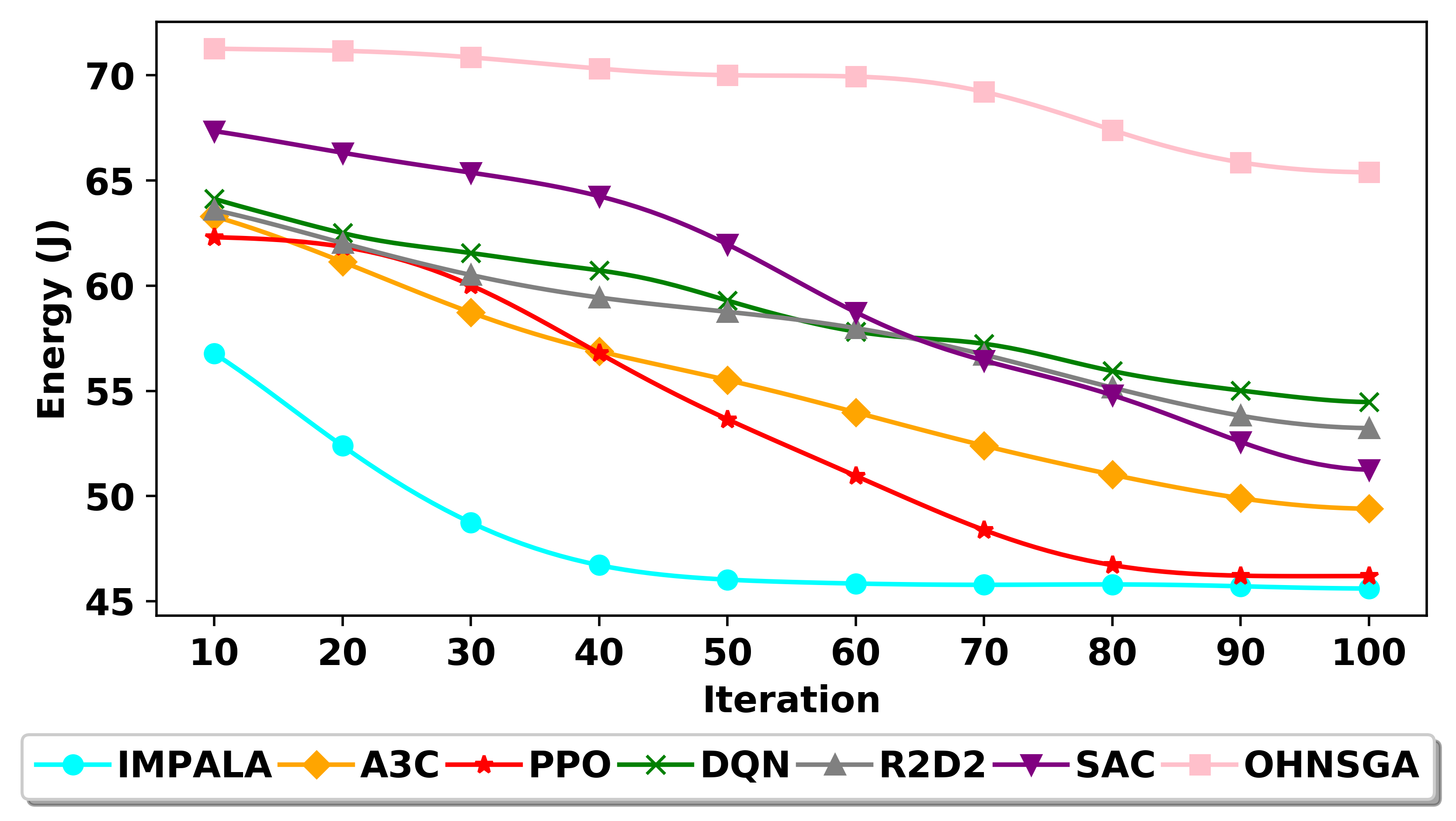}
  \caption{Energy consumption}
  \label{fig:ent}
\end{subfigure}
\begin{subfigure}{0.33\textwidth}
  \centering
  \includegraphics[width=\linewidth]{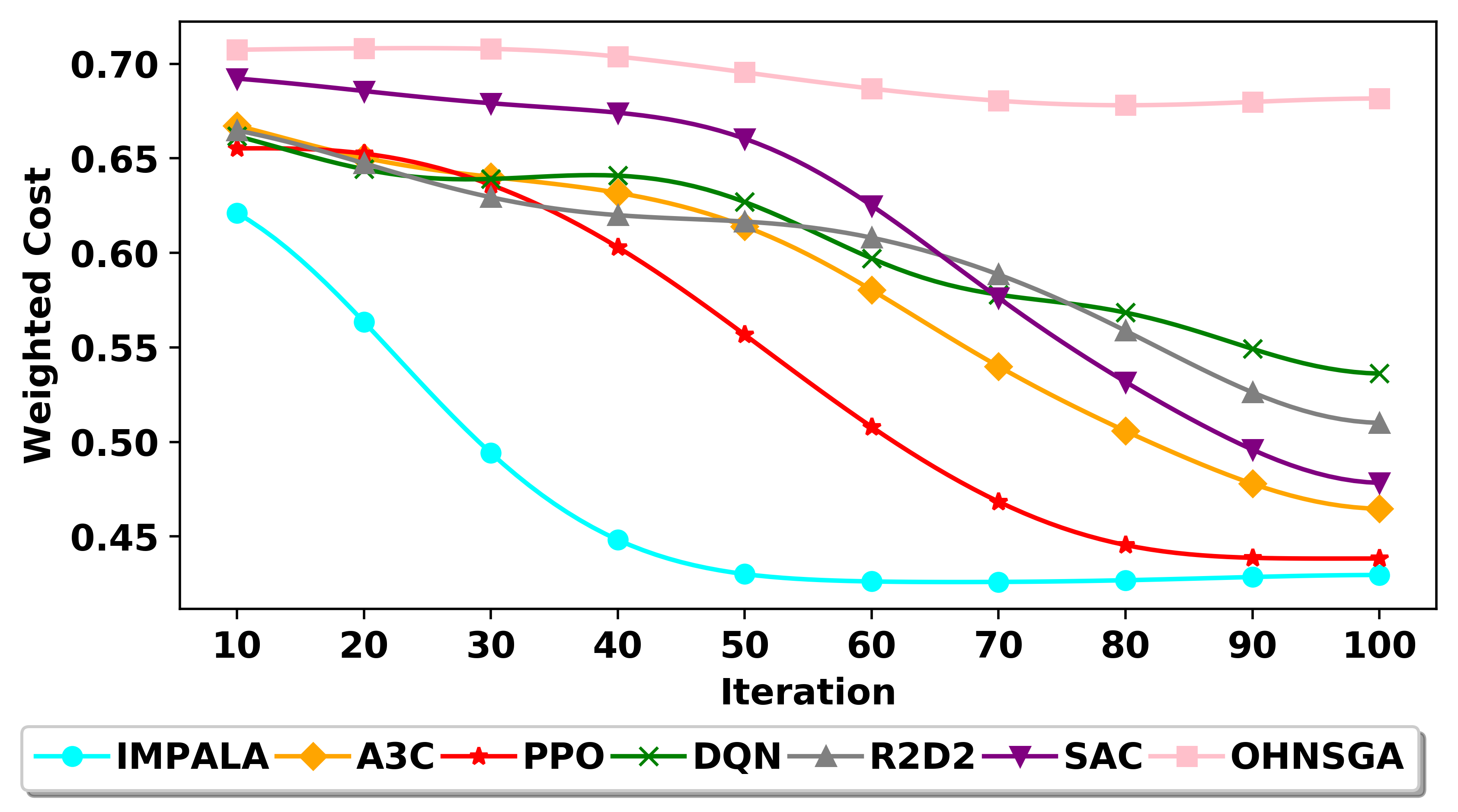}
  \caption{Weighted cost}
  \label{fig:wct}
\end{subfigure}
\caption{Convergence performance comparison of scheduling techniques during training phase}
\label{fig:train}
\end{figure*}

\begin{figure*}[pos=t]
\begin{subfigure}{0.33\textwidth}
  \centering
  \includegraphics[width=\linewidth]{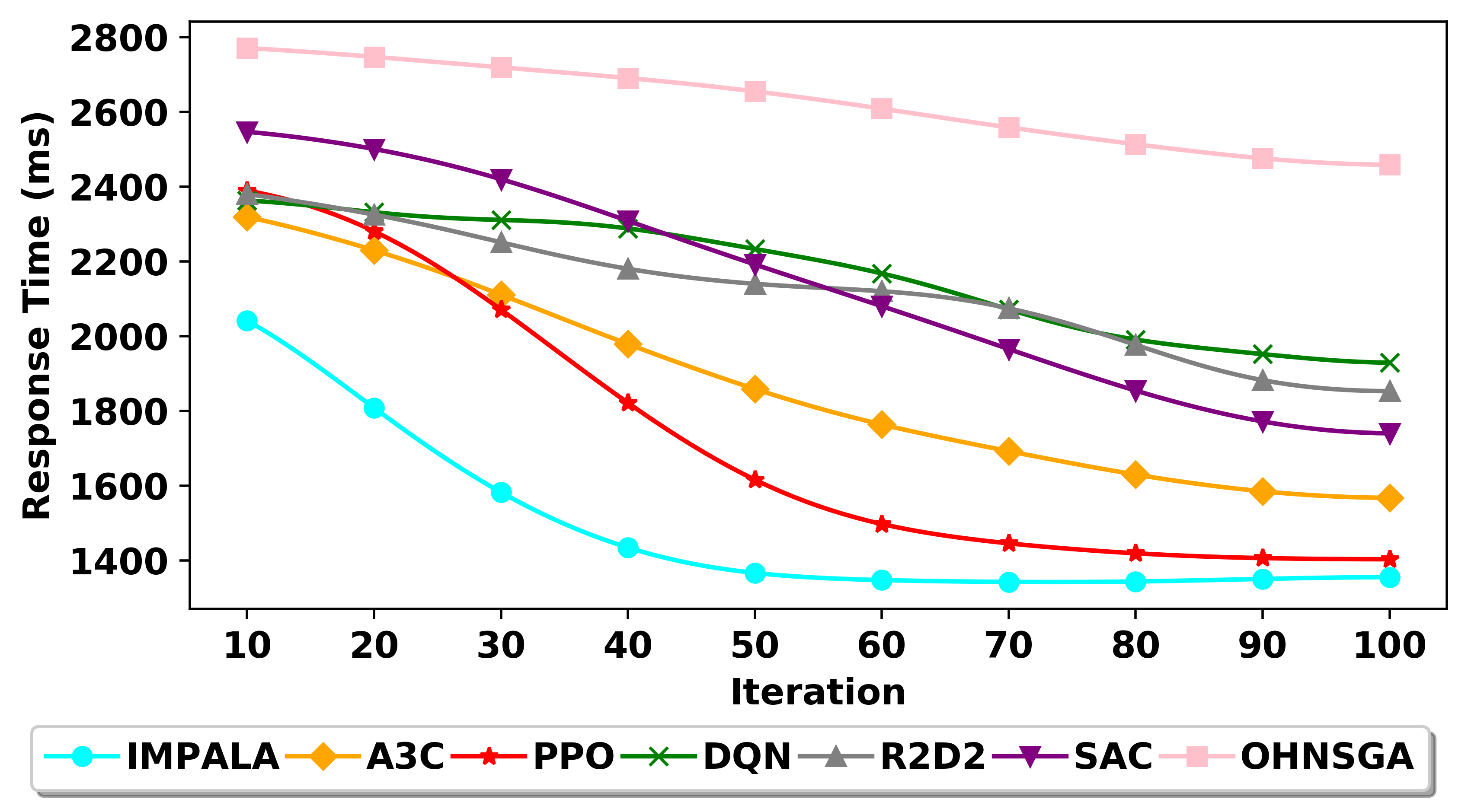}
  \caption{Response time}
  \label{fig:rtt}
\end{subfigure}%
\begin{subfigure}{0.33\textwidth}
  \centering
  \includegraphics[width=\linewidth]{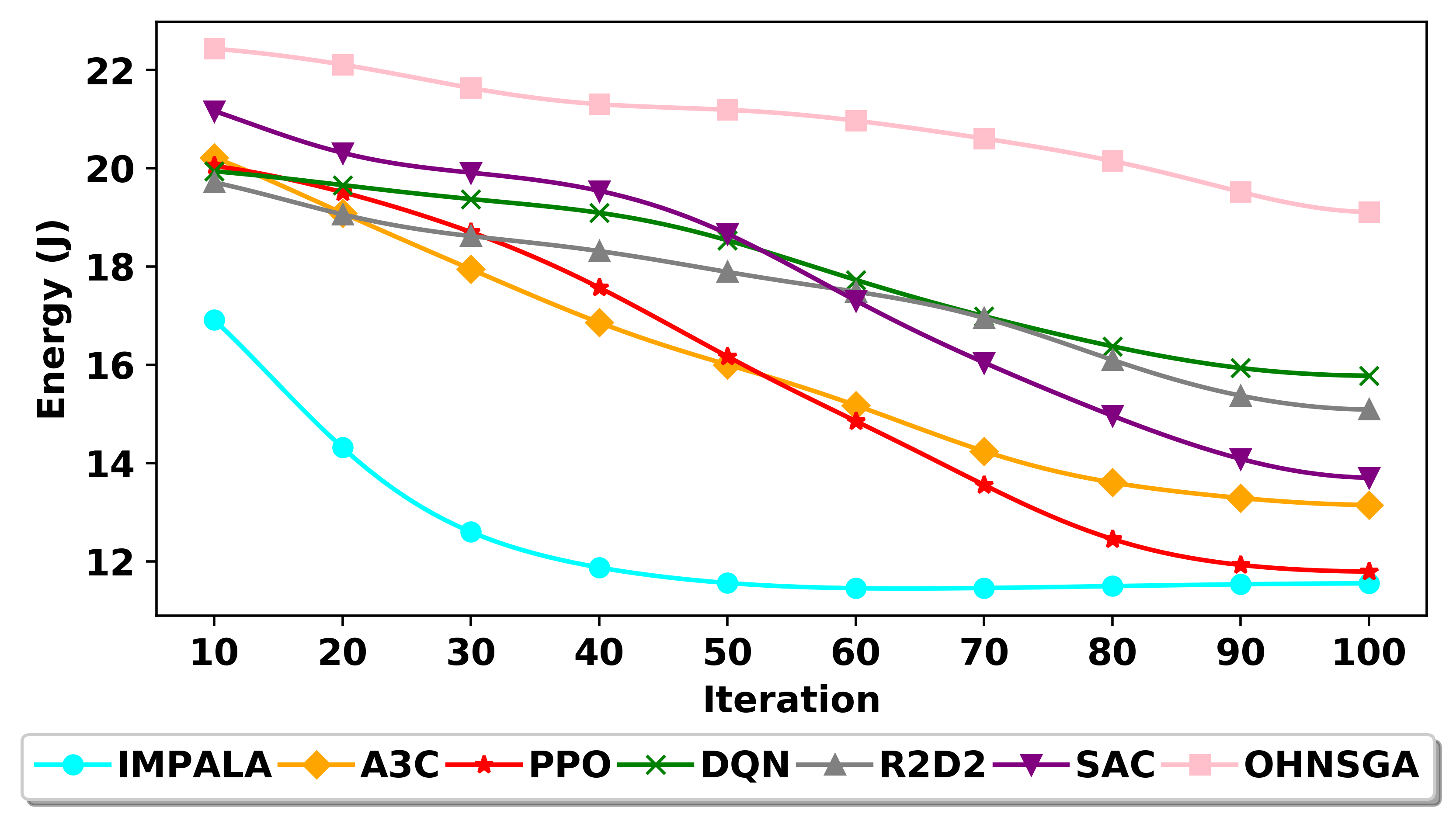}
  \caption{Energy consumption}
  \label{fig:ent}
\end{subfigure}
\begin{subfigure}{0.33\textwidth}
  \centering
  \includegraphics[width=\linewidth]{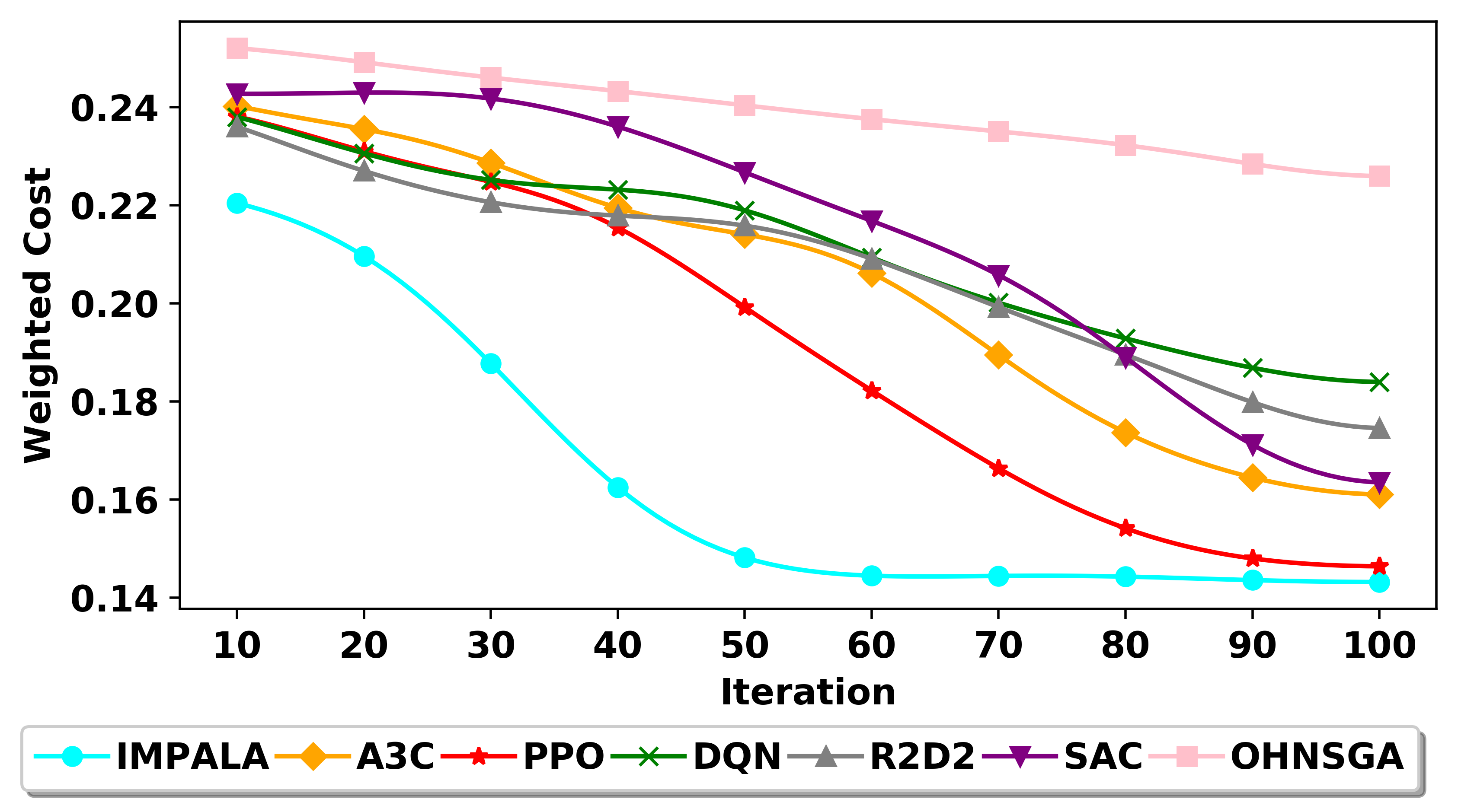}
  \caption{Weighted cost}
  \label{fig:wct}
\end{subfigure}
\caption{Convergence performance comparison of scheduling techniques during evaluation phase}
\label{fig:eva}
\end{figure*}

In this experiment, we evaluate and compare the convergence performance of various DRL techniques in ReinFog with OHNSGA during both training and evaluation phases. The comparison is conducted across three metrics: response time, energy consumption, and weighted cost, as shown in Fig. \ref{fig:train} and Fig. \ref{fig:eva}. During the training phase, the video resolution is set to 480p, while in the evaluation phase, it is reduced to 240p to test the adaptability of different techniques. In the training phases, across all three metrics, IMPALA consistently demonstrates superior performance, achieving the fastest convergence (around iteration 40-50) and the lowest final values. PPO emerges as the second-best performer, followed by A3C with moderate performance. The imported techniques, R2D2 and SAC, converge more slowly, while DQN demonstrates relatively stable but the slowest convergence among all DRL techniques. OHNSGA, the meta-heuristic baseline from FogBus2, consistently performs the worst, showing minimal improvement over iterations and the highest final values. The evaluation phase maintains similar performance patterns but with overall lower metric values due to the reduced video resolution. In the evaluation phase, ReinFog's DRL techniques keep achieving significant improvements over OHNSGA, with reductions of up to 45\% in response time, 39\% in energy consumption, and 37\% in weighted cost. The results demonstrate the robustness and generalization capability of ReinFog's DRL techniques, particularly IMPALA, in adapting to varying workload conditions. The significant performance gap between ReinFog's DRL techniques and FogBus2's OHNSGA validates the effectiveness of DRL-based approaches in dynamic IoT application scheduling scenarios.

\subsubsection{Scheduling Techniques Overhead Analysis}
\begin{figure}[pos=t]
\includegraphics[width=\linewidth]{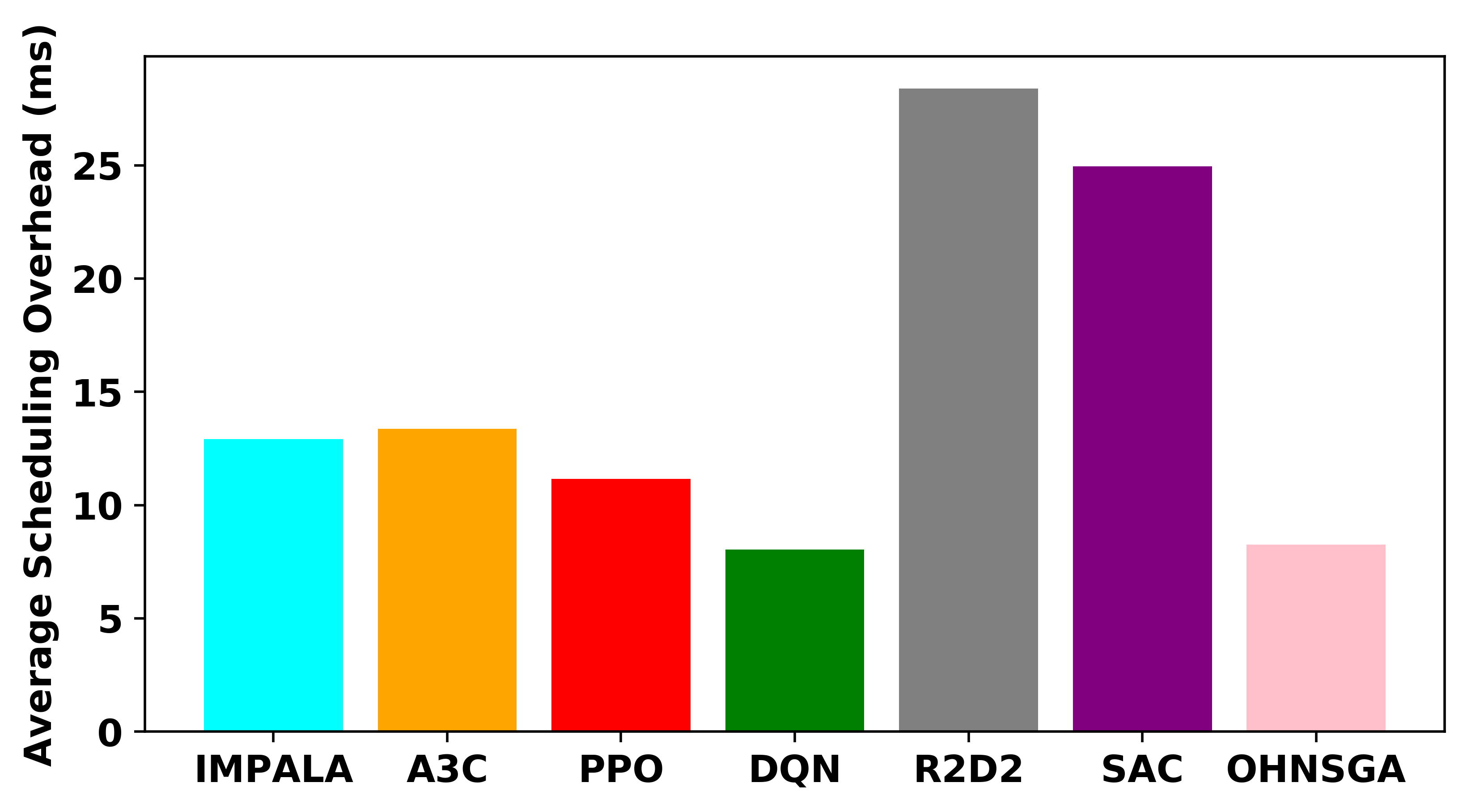}
\caption{Average scheduling overhead comparison across different scheduling techniques}\label{fig:aso}
\end{figure}

In this experiment, we compare the average scheduling overhead of different techniques in ReinFog, against OHNSGA, as shown in Fig. \ref{fig:aso}. Among ReinFog's native DRL techniques, DQN shows the lowest overhead at approximately 8ms, followed by PPO (11ms), while IMPALA and A3C both demonstrate an overhead of around 13ms. The imported techniques demonstrate higher overhead, with R2D2 at around 28ms and SAC at 25ms. FogBus2's OHNSGA exhibits the lowest overhead at 8ms. The results show that native integrated DRL techniques (IMPALA, A3C, PPO, DQN) consistently demonstrate lower overhead than external imported ones (R2D2, SAC), and centralized techniques (DQN, PPO) generally incur less overhead than distributed ones (IMPALA, A3C). While OHNSGA shows the lowest overhead, the superior convergence performance of ReinFog's DRL techniques justifies their moderate scheduling overhead.

\subsubsection{Scheduling Techniques Scalability Analysis}
\begin{figure}[pos=t]
\includegraphics[width=\linewidth]{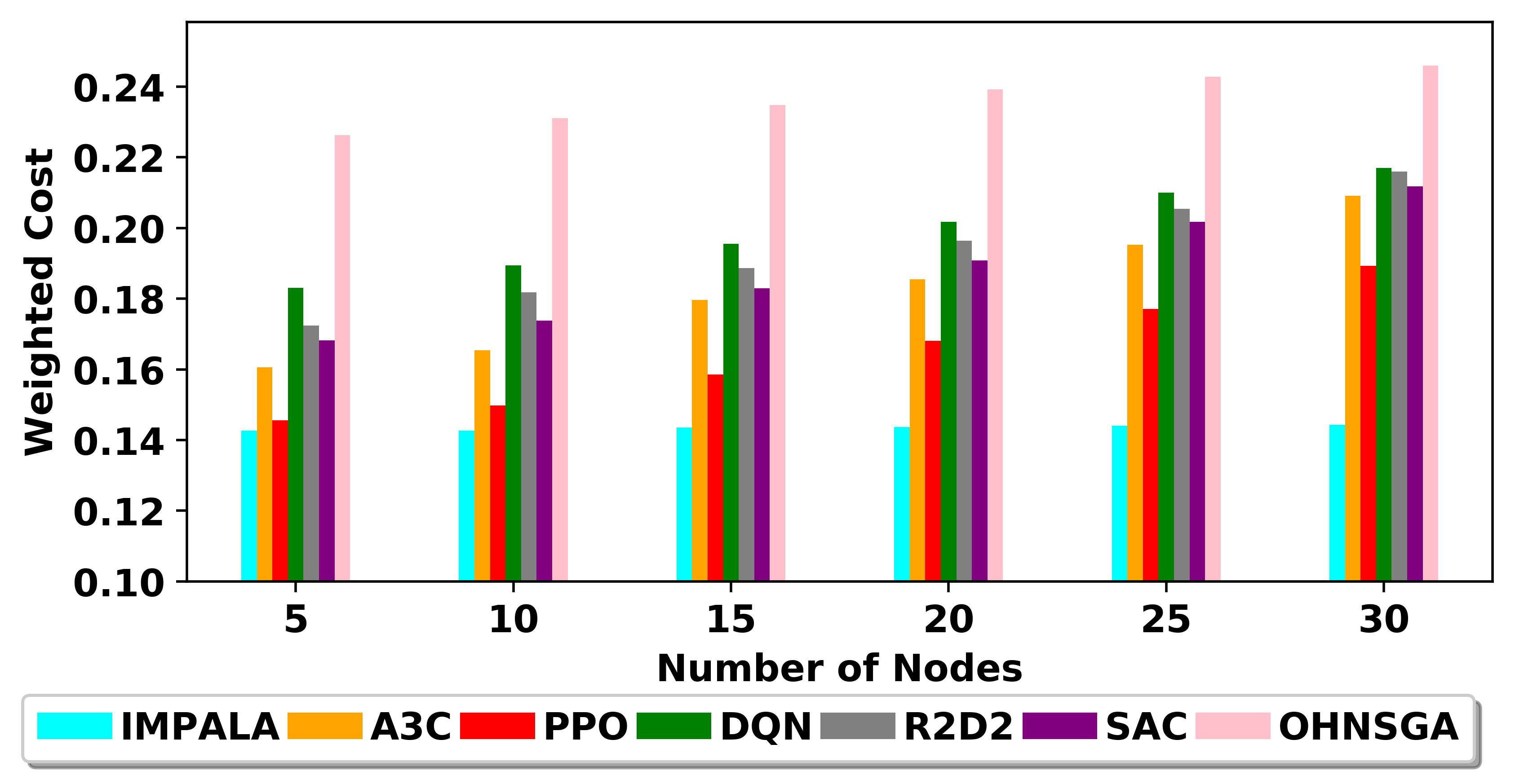}
\caption{Impact of number of nodes on scheduling performance across different scheduling techniques}\label{fig:sct}
\end{figure}

In this experiment, we evaluate and compare the scalability of different scheduling techniques by varying the number of nodes from 5 to 30. As the response time and energy consumption metrics show similar trends, we present only the weighted cost results here. As shown in Fig. \ref{fig:sct}, IMPALA consistently achieves the lowest and most stable weighted cost (around 0.14) across all node configurations, showing excellent scalability. Other DRL techniques, including A3C, PPO, DQN, R2D2, and SAC, demonstrate moderate increases in weighted cost as the number of nodes grows. While their relative performance shows some variations across different scales, they all maintain significantly better performance than OHNSGA. Specifically, A3C and PPO show better scalability with lower weighted costs, SAC maintains intermediate performance, while DQN and R2D2 exhibit relatively higher weighted costs as the system scales up. In contrast, OHNSGA exhibits the highest weighted cost across all scales, maintaining its weighted cost between 0.22 and 0.25 as the system scales up. This analysis clearly demonstrates that while some DRL techniques in ReinFog exhibit performance fluctuations, overall they demonstrate good scalability in scheduling IoT applications across different nodes. In contrast, traditional meta-heuristic methods like OHNSGA perform poorly across all environment scales. 

\subsubsection{Scheduling Techniques Sustainability Analysis}
\begin{figure}[pos=t]
\includegraphics[width=\linewidth]{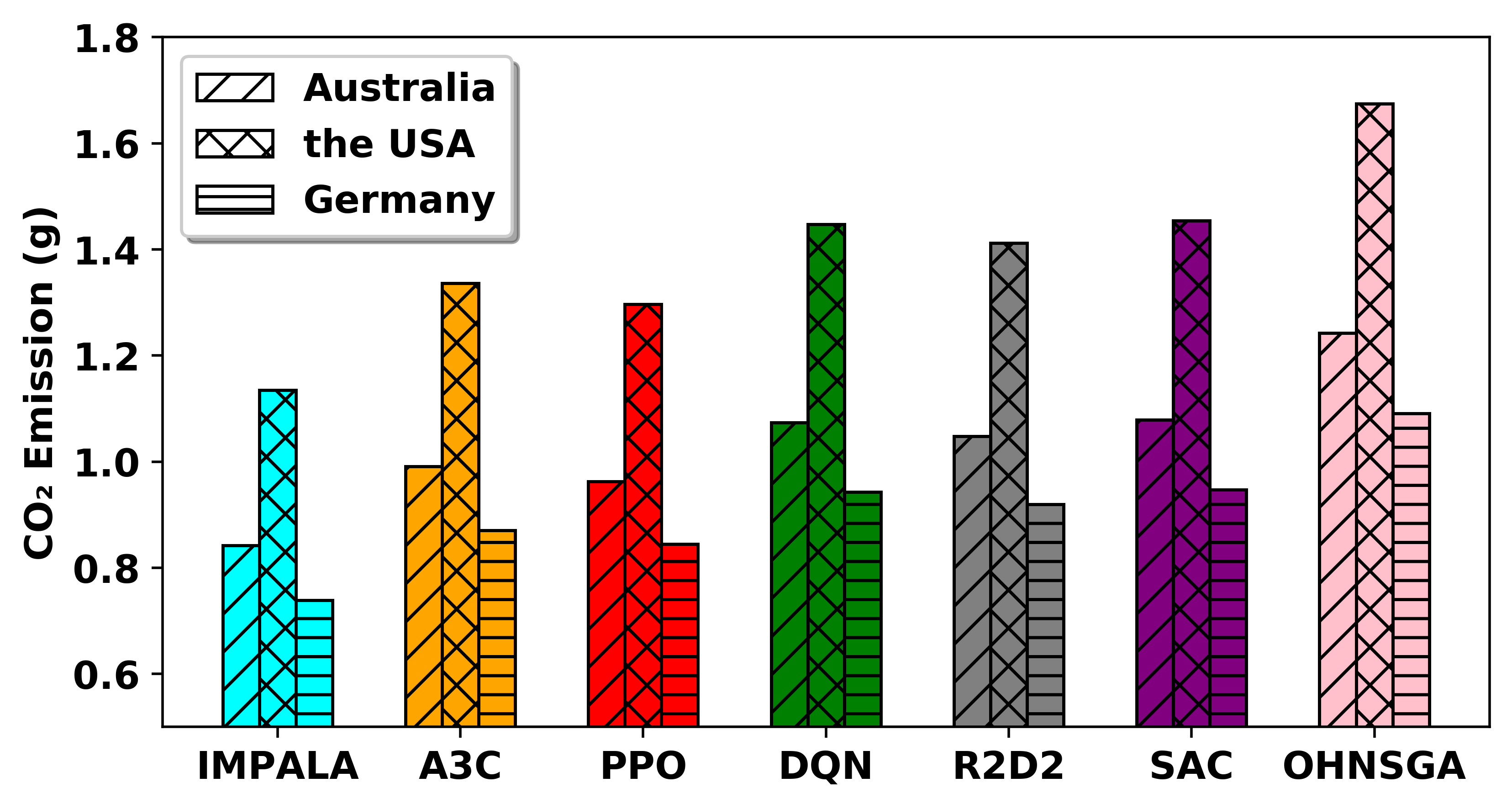}
\caption{Hourly \(\text{CO}_2\) emissions comparison of scheduling techniques across different regions}\label{fig:coe}
\end{figure}

In this experiment, we evaluate the environmental sustainability of different scheduling techniques by comparing the \(\text{CO}_2\) emissions when processing IoT applications for one hour. We estimate the \(\text{CO}_2\) emissions in Australia, the USA, and Germany following the same approach described in Section \ref{carbon}. The results are shown in Fig. \ref{fig:coe}. In general, all DRL techniques demonstrate lower \(\text{CO}_2\) emissions compared to OHNSGA. Among the DRL techniques, IMPALA shows the lowest emissions across all regions (0.84g in Australia, 1.14g in the USA, and 0.74g in Germany), while DQN exhibits the highest emissions among DRL techniques (1.08g in Australia, 1.45g in the USA, and 0.94g in Germany). OHNSGA produces significantly higher emissions (1.25g in Australia, 1.68g in the USA, and 1.10g in Germany), approximately 50\% more than IMPALA. Notably, all techniques show consistently higher emissions in the USA while Germany demonstrates the lowest emissions among the three regions. This pattern can be attributed to differences in regional electricity generation patterns and their corresponding carbon intensities. These results demonstrate that the DRL techniques in ReinFog can better contribute to environmental sustainability compared to traditional meta-heuristic methods.

\subsection{DRL Component Placement Algorithm Analysis}
In this section, we evaluate the performance of our proposed MADCP in comparison with other DRL component placement algorithms, including GA, FA, PSO, and a random placement method. Given that IMPALA demonstrates superior performance in our previous convergence analysis, we employ it with various DRL component placement algorithms to assess their impact on overall performance. This study focuses on analyzing the effects of various DRL component placement algorithms on the convergence rate and scheduling overhead of IMPALA.

\subsubsection{DRL Component Placement Algorithm Convergence Analysis}
To assess the impact of different DRL component placement algorithms on IMPALA's performance, we conducted a convergence analysis across the three metrics: response time, energy consumption, and weighted cost. The results are demonstrated in Figure \ref{fig:ale}. MADCP consistently outperforms other DRL component placement algorithms across all metrics, achieving the fastest convergence rates, typically stabilizing after 50-60 iterations. PSO and GA follow closely, performing similarly and surpassing FA. Although the FA method converges slightly more slowly, it still outperforms the random placement method, which shows the poorest performance across all metrics and only stabilizes after 90 iterations. Notably, MADCP accelerates the convergence rate by up to 38\% compared to random placement method. These results highlight MADCP's effectiveness in providing a strong starting point for DRL techniques. Its superior performance stems from combining the strengths of GA, FA, and PSO, leading to more efficient exploration of the solution space. This analysis also underscores the importance of intelligent placement of DRL components in DRL-based resource management frameworks.
\begin{figure*}[pos=t]
\begin{subfigure}{0.33\textwidth}
  \centering
  \includegraphics[width=\linewidth]{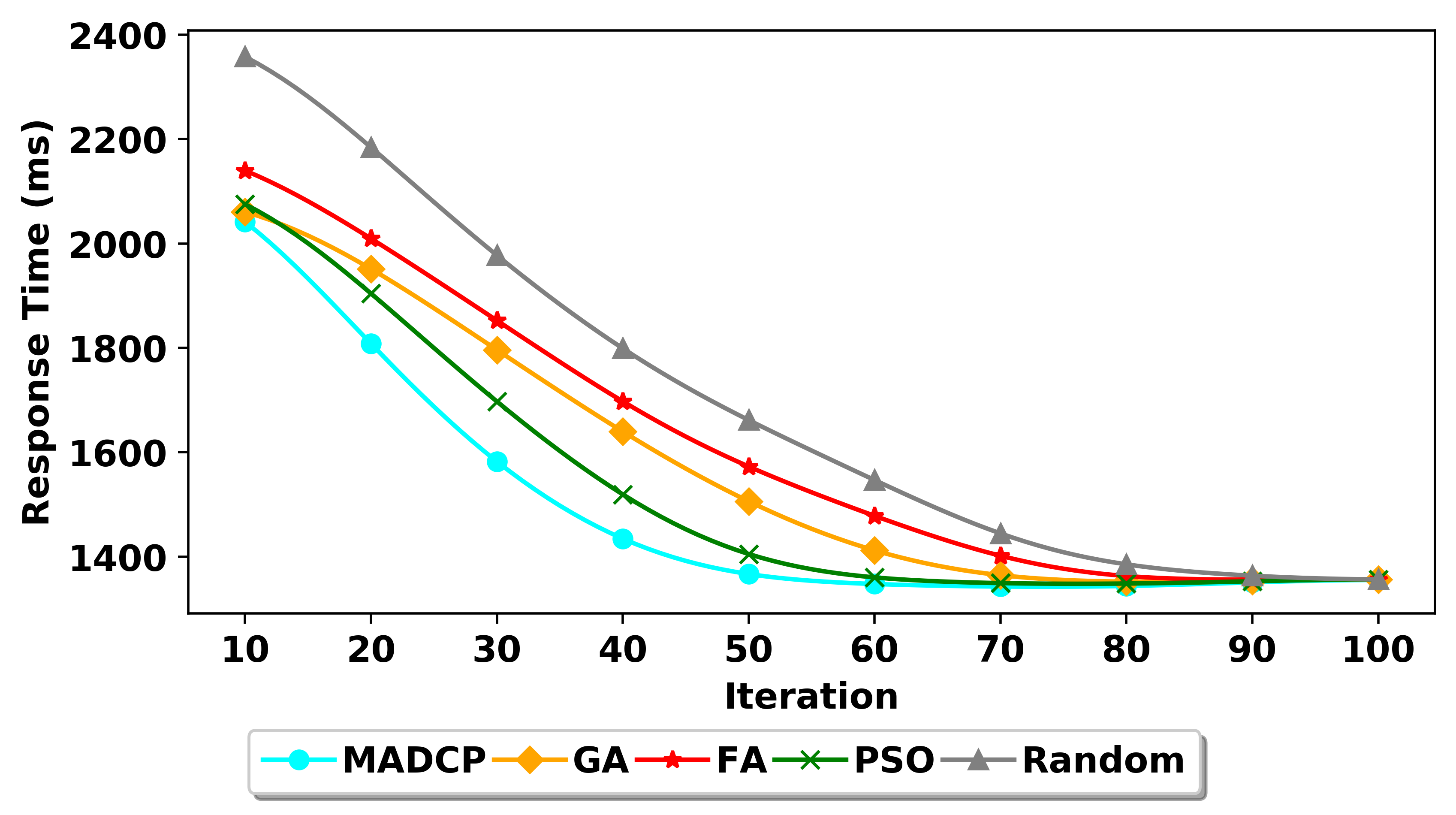}
  \caption{Response time}
  \label{fig:alt}
\end{subfigure}%
\begin{subfigure}{0.33\textwidth}
  \centering
  \includegraphics[width=\linewidth]{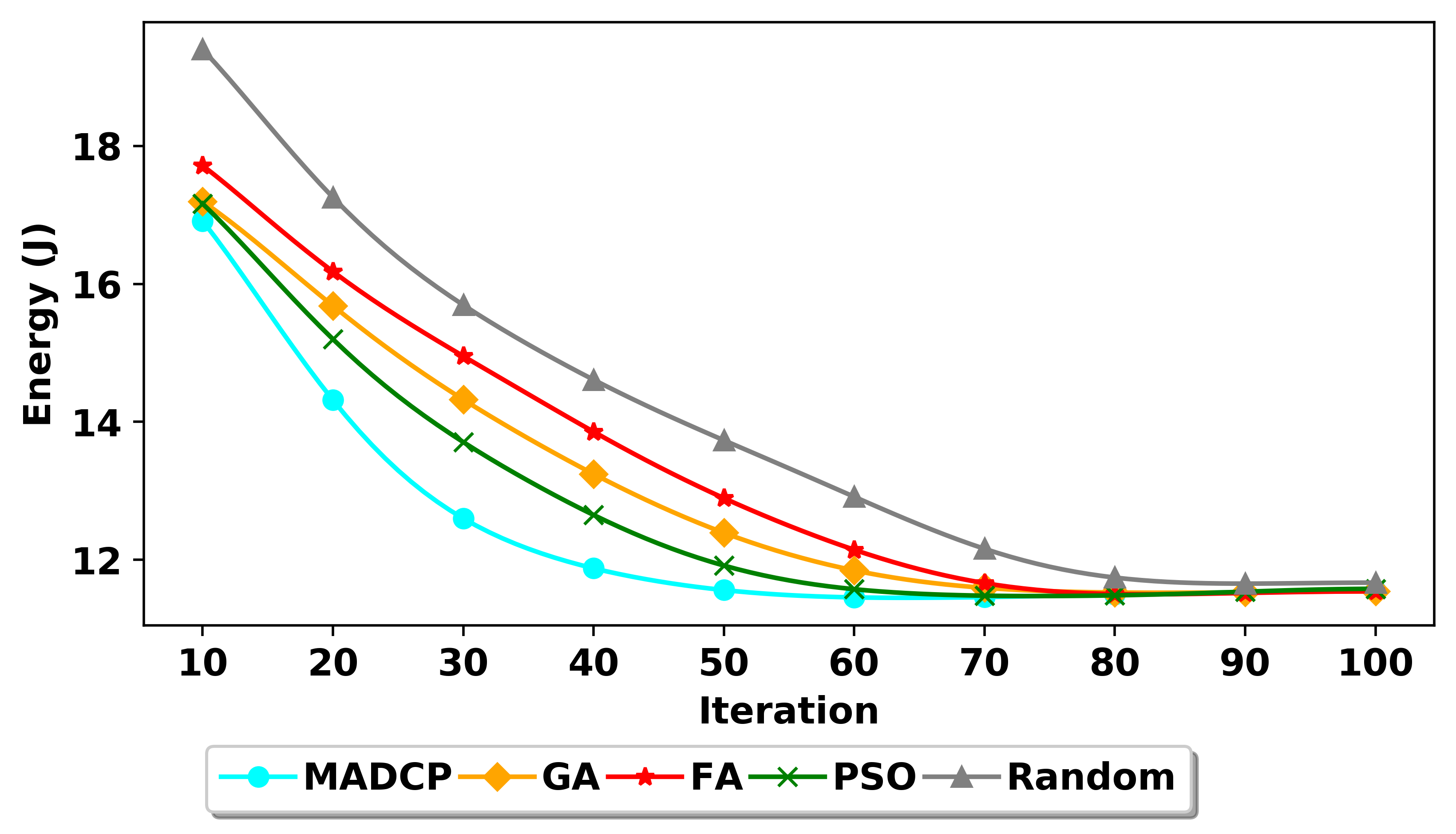}
  \caption{Energy consumption}
  \label{fig:ale}
\end{subfigure}
\begin{subfigure}{0.33\textwidth}
  \centering
  \includegraphics[width=\linewidth]{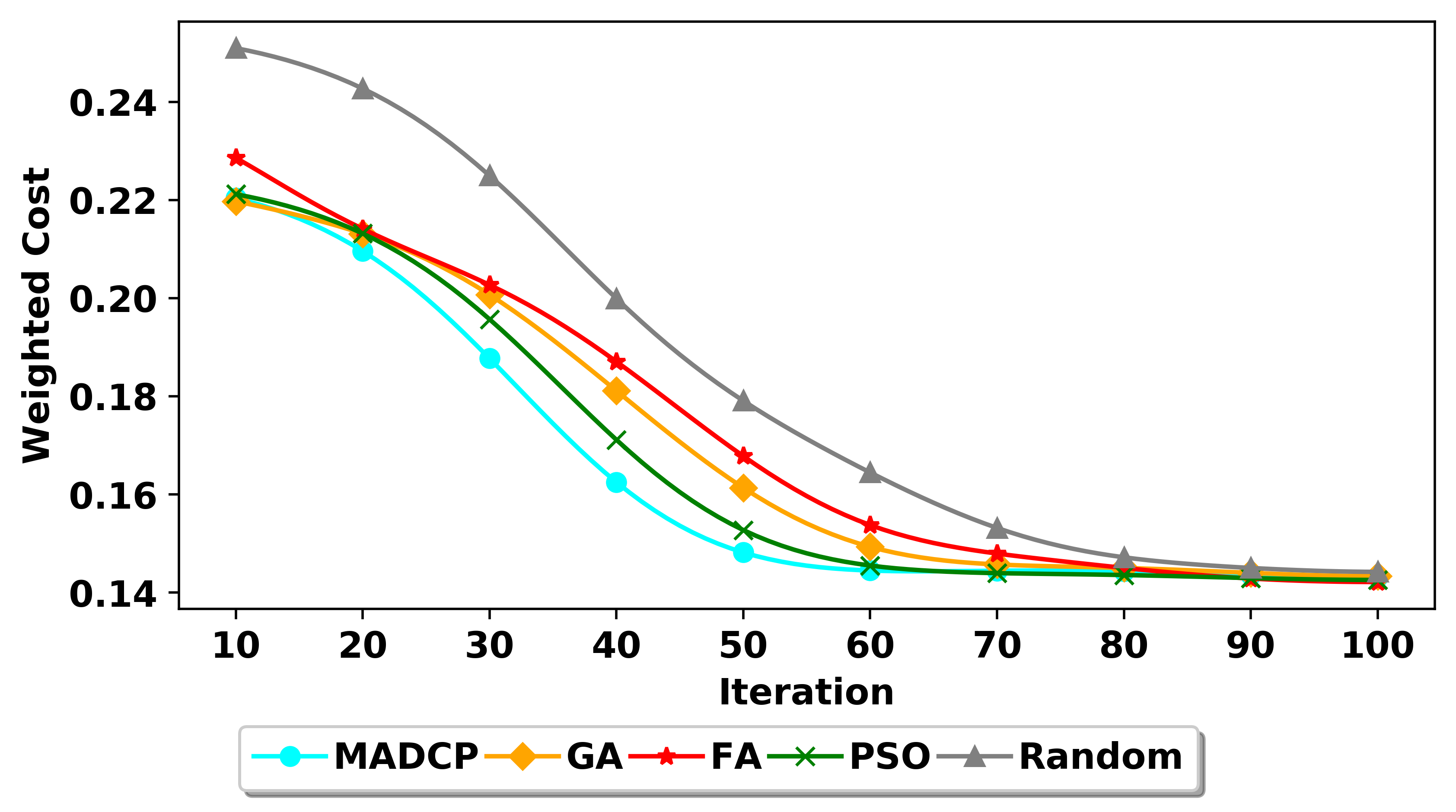}
  \caption{Weighted cost}
  \label{fig:alw}
\end{subfigure}
\caption{Impact of different DRL component placement algorithms on IMPALA's convergence performance}
\label{fig:ale}
\end{figure*}

\subsubsection{DRL Component Placement Algorithm Overhead Analysis}
\begin{figure}[pos=t]
\includegraphics[width=\linewidth]{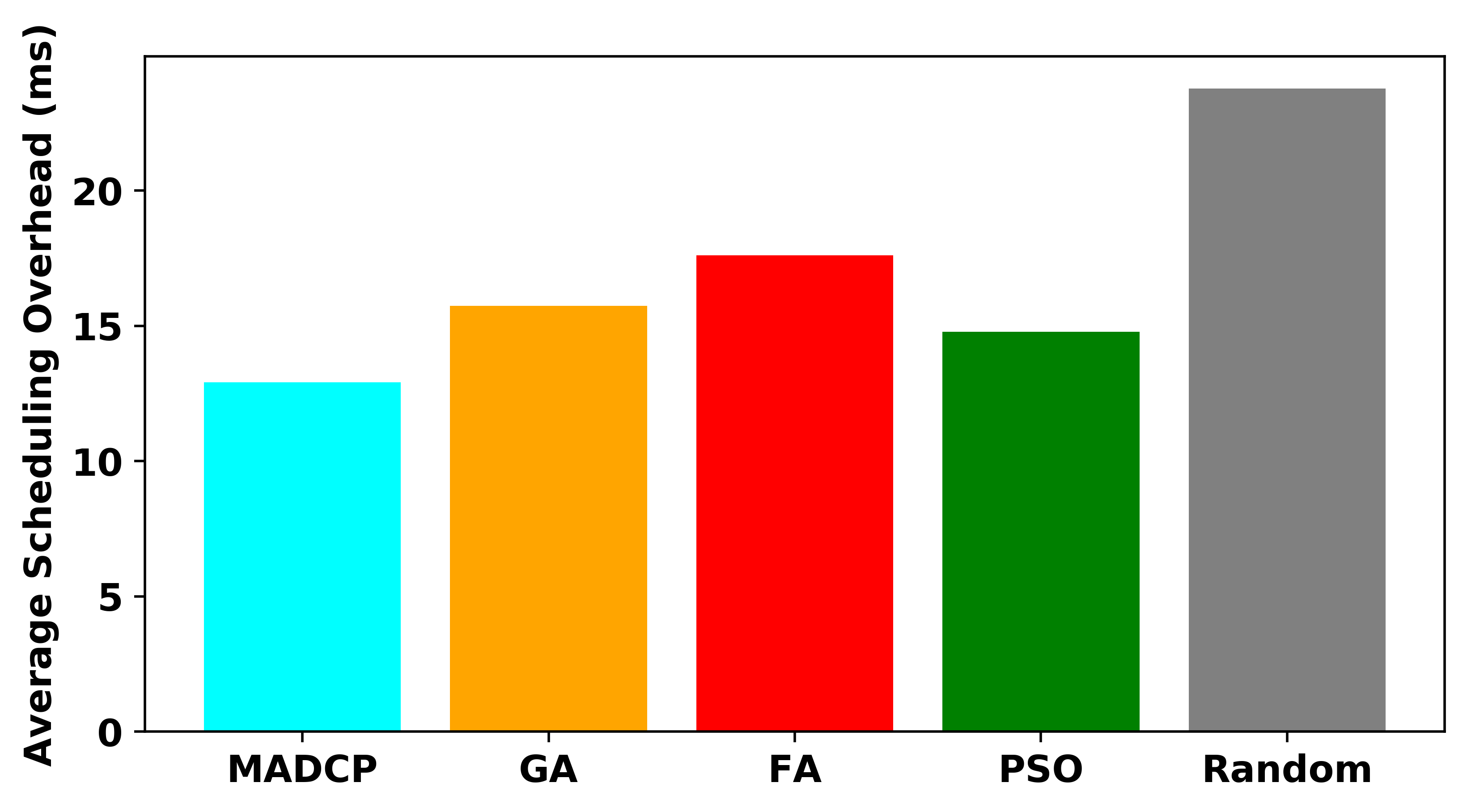}
\caption{Impact of different DRL component placement algorithms on IMPALA's average scheduling overhead}\label{fig:alo}
\end{figure}

In this experiment, we evaluate and compare the average scheduling overhead of IMPALA when employed with different DRL component placement algorithms. As shown in Fig. \ref{fig:alo}, MADCP demonstrates the lowest average overhead at approximately 13ms, followed by PSO at 15ms, GA at 16ms, and FA at 18ms. The random placement method shows the highest overhead at about 24ms, nearly twice that of MADCP. These results highlight MADCP's effectiveness in reducing computational overhead when used to place DRL components in the ReinFog framework.

\section{Conclusions and Future Work}
This paper proposed ReinFog, a novel framework leveraging DRL mechanisms and techniques for adaptive resource management in edge/fog and cloud computing environments. ReinFog addresses the challenge of efficiently scheduling heterogeneous IoT applications across diverse computing resources through its modular and extensible DRL components. It offers capabilities to support centralized and distributed DRL techniques and allows integration of both native and library-based DRL techniques. It features customizable deployment configurations, allowing users to flexibly configure DRL Learners and Workers based on system requirements. Additionally, it incorporates MADCP, an efficient DRL component placement algorithm that dynamically optimizes the allocation of DRL Learners and Workers, enhancing DRL-based scheduling techniques performance in distributed environments. Our extensive experiments demonstrate that ReinFog is a lightweight and scalable framework capable of effectively scheduling IoT applications under diverse optimization objectives.
\par
As part of our future work, we will extend ReinFog by integrating security and privacy mechanisms to enable secure DRL parameter-sharing and experience-sharing approaches. The framework's modular design allows seamless integration of security protocols at all levels, ensuring that security features can be embedded for secure communication and data handling without affecting core functionality or system efficiency. Additionally, we will integrate more recent DRL techniques into ReinFog, which will allow the framework to remain at the forefront of cutting-edge research. This integration will enable researchers to leverage advanced algorithms, providing a flexible platform for experimenting with and refining novel DRL techniques for diverse research. Furthermore, we plan to investigate techniques to improve system resilience against hardware and software failures, including fault-tolerant scheduling and adaptive component redistribution across heterogeneous nodes.

\bibliographystyle{elsarticle-num}

\bibliography{cas-refs}

\end{document}